%
%
%
\def\unredoffs{} \def\redoffs{\voffset=-.31truein\hoffset=-.48truein}
\def\speclscape{}
%
%
%
%
%
\newbox\leftpage \newdimen\fullhsize \newdimen\hstitle \newdimen\hsbody
\tolerance=1000\hfuzz=2pt
\catcode`\@=11 
\ifx\hyperdef\UNd@FiNeD\def\hyperdef#1#2#3#4{#4}\def\hyperref#1#2#3#4{#4}\fi
\def\bigans{b }
\def\answ{b }
%
\ifx\answ\bigans\message{(This will come out unreduced.}
\magnification=1200\unredoffs\baselineskip=16pt plus 2pt minus 1pt
\hsbody=\hsize \hstitle=\hsize 
\else\message{(This will be reduced.} \let\l@r=L
\magnification=1000\baselineskip=16pt plus 2pt minus 1pt \vsize=7truein
\redoffs \hstitle=8truein\hsbody=4.75truein\fullhsize=10truein\hsize=\hsbody
\output={\ifnum\pageno=0 
  \shipout\vbox{\speclscape{\hsize\fullhsize\makeheadline}
    \hbox to \fullhsize{\hfill\pagebody\hfill}}\advancepageno
  \else
  \almostshipout{\leftline{\vbox{\pagebody\makefootline}}}\advancepageno
  \fi}
\def\almostshipout#1{\if L\l@r \count1=1 \message{[\the\count0.\the\count1]}
      \global\setbox\leftpage=#1 \global\let\l@r=R
 \else \count1=2
  \shipout\vbox{\speclscape{\hsize\fullhsize\makeheadline}
      \hbox to\fullhsize{\box\leftpage\hfil#1}}  \global\let\l@r=L\fi}
\fi
%
\newcount\yearltd\yearltd=\year\advance\yearltd by -2000

\def\Title#1#2{\nopagenumbers\abstractfont\hsize=\hstitle\rightline{#1}%
\vskip 1in\centerline{\titlefont #2}\abstractfont\vskip .5in\pageno=0}
\def\Date#1{\vfill\leftline{#1}\tenpoint\supereject\global\hsize=\hsbody%
\footline={\hss\tenrm\hyperdef\hypernoname{page}\folio\folio\hss}}%
%

\def\draftmode{\message{ DRAFTMODE }\def\draftdate{{\rm preliminary draft:
\number\month/\number\day/\number\yearltd\ \ \hourmin}}%
\headline={\hfil\draftdate}\writelabels\baselineskip=20pt plus 2pt minus 2pt
 {\count255=\time\divide\count255 by 60 \xdef\hourmin{\number\count255}
  \multiply\count255 by-60\advance\count255 by\time
  \xdef\hourmin{\hourmin:\ifnum\count255<10 0\fi\the\count255}}}
\def\nolabels{\def\wrlabeL##1{}\def\eqlabeL##1{}\def\reflabeL##1{}}
\def\writelabels{\def\wrlabeL##1{\leavevmode\vadjust{\rlap{\smash%
{\line{{\escapechar=` \hfill\rlap{\sevenrm\hskip.03in\string##1}}}}}}}%
\def\eqlabeL##1{{\escapechar-1\rlap{\sevenrm\hskip.05in\string##1}}}%
\def\reflabeL##1{\noexpand\llap{\noexpand\sevenrm\string\string\string##1}}}
\nolabels
%
\global\newcount\secno \global\secno=0
\global\newcount\meqno \global\meqno=1
\def\s@csym{}
\def\newsec#1{\global\advance\secno by1%
{\toks0{#1}\message{(\the\secno. \the\toks0)}}%
\global\subsecno=0\eqnres@t\let\s@csym\secsym\xdef\secn@m{\the\secno}\noindent
{\bf\hyperdef\hypernoname{section}{\the\secno}{\the\secno.} #1}%
\writetoca{{\string\hyperref{}{section}{\the\secno}{\the\secno.}} {#1}}%
\par\nobreak\medskip\nobreak}
\def\eqnres@t{\xdef\secsym{\the\secno.}\global\meqno=1\bigbreak\bigskip}
\def\sequentialequations{\def\eqnres@t{\bigbreak}}\xdef\secsym{}
\global\newcount\subsecno \global\subsecno=0
\def\subsec#1{\global\advance\subsecno by1%
{\toks0{#1}\message{(\s@csym\the\subsecno. \the\toks0)}}%
\ifnum\lastpenalty>9000\else\bigbreak\fi
\noindent{\it\hyperdef\hypernoname{subsection}{\secn@m.\the\subsecno}%
{\secn@m.\the\subsecno.} #1}\writetoca{\string\quad
{\string\hyperref{}{subsection}{\secn@m.\the\subsecno}{\secn@m.\the\subsecno.}}
{#1}}\par\nobreak\medskip\nobreak}
\def\appendix#1#2{\global\meqno=1\global\subsecno=0\xdef\secsym{\hbox{#1.}}%
\bigbreak\bigskip\noindent{\bf Appendix \hyperdef\hypernoname{appendix}{#1}%
{#1.} #2}{\toks0{(#1. #2)}\message{\the\toks0}}%
\xdef\s@csym{#1.}\xdef\secn@m{#1}%
\writetoca{\string\hyperref{}{appendix}{#1}{Appendix {#1.}} {#2}}%
\par\nobreak\medskip\nobreak}
%
%
\def\checkm@de#1#2{\ifmmode{\def\f@rst##1{##1}\hyperdef\hypernoname{equation}%
{#1}{#2}}\else\hyperref{}{equation}{#1}{#2}\fi}
\def\eqnn#1{\DefWarn#1\xdef #1{(\noexpand\relax\noexpand\checkm@de%
{\s@csym\the\meqno}{\secsym\the\meqno})}%
\wrlabeL#1\writedef{#1\leftbracket#1}\global\advance\meqno by1}
\def\f@rst#1{\c@t#1a\em@ark}\def\c@t#1#2\em@ark{#1}
\def\eqna#1{\DefWarn#1\wrlabeL{#1$\{\}$}%
\xdef #1##1{(\noexpand\relax\noexpand\checkm@de%
{\s@csym\the\meqno\noexpand\f@rst{##1}}{\hbox{$\secsym\the\meqno##1$}})}
\writedef{#1\numbersign1\leftbracket#1{\numbersign1}}\global\advance\meqno by1}
\def\eqn#1#2{\DefWarn#1%
\xdef #1{(\noexpand\hyperref{}{equation}{\s@csym\the\meqno}%
{\secsym\the\meqno})}$$#2\eqno(\hyperdef\hypernoname{equation}%
{\s@csym\the\meqno}{\secsym\the\meqno})\eqlabeL#1$$%
\writedef{#1\leftbracket#1}\global\advance\meqno by1}
\def\xeqn{\expandafter\xe@n}\def\xe@n(#1){#1}
\def\xeqna#1{\expandafter\xe@n#1}
\def\eqns#1{(\e@ns #1{\hbox{}})}
\def\e@ns#1{\ifx\UNd@FiNeD#1\message{eqnlabel \string#1 is undefined.}%
\xdef#1{(?.?)}\fi{\let\hyperref=\relax\xdef\next{#1}}%
\ifx\next\em@rk\def\next{}\else%
\ifx\next#1\xeqn#1\else\def\n@xt{#1}\ifx\n@xt\next#1\else\xeqna#1\fi
\fi\let\next=\e@ns\fi\next}

\def\DefWarn#1{\ifx\UNd@FiNeD#1\else
\immediate\write16{*** WARNING: the label \string#1 is already defined ***}\fi}
%
\newskip\footskip\footskip14pt plus 1pt minus 1pt 
\def\footnotefont{\ninepoint}\def\f@t#1{\footnotefont #1\@foot}
\def\f@@t{\baselineskip\footskip\bgroup\footnotefont\aftergroup\@foot\let\next}
\setbox\strutbox=\hbox{\vrule height9.5pt depth4.5pt width0pt}
\global\newcount\ftno \global\ftno=0
\def\foot{\global\advance\ftno by1\def\foot@rg{\hyperref{}{footnote}%
{\the\ftno}{\the\ftno}\xdef\foot@rg{\noexpand\hyperdef\noexpand\hypernoname%
{footnote}{\the\ftno}{\the\ftno}}}\footnote{$^{\foot@rg}$}}
%
\newwrite\ftfile
\def\footend{\def\foot{\global\advance\ftno by1\chardef\wfile=\ftfile
\hyperref{}{footnote}{\the\ftno}{$^{\the\ftno}$}%
\ifnum\ftno=1\immediate\openout\ftfile=\jobname.fts\fi%
\immediate\write\ftfile{\noexpand\smallskip%
\noexpand\item{\noexpand\hyperdef\noexpand\hypernoname{footnote}
{\the\ftno}{f\the\ftno}:\ }\pctsign}\findarg}%
\def\footatend{\vfill\eject\immediate\closeout\ftfile{\parindent=20pt
\centerline{\bf Footnotes}\nobreak\bigskip\input \jobname.fts }}}
\def\footatend{}
%
%
\global\newcount\refno \global\refno=1
\newwrite\rfile
\def\ref{[\hyperref{}{reference}{\the\refno}{\the\refno}]\nref}
\def\nref#1{\DefWarn#1%
\xdef#1{[\noexpand\hyperref{}{reference}{\the\refno}{\the\refno}]}%
\writedef{#1\leftbracket#1}%
\ifnum\refno=1\immediate\openout\rfile=\jobname.refs\fi
\chardef\wfile=\rfile\immediate\write\rfile{\noexpand\item{[\noexpand\hyperdef%
\noexpand\hypernoname{reference}{\the\refno}{\the\refno}]\ }%
\reflabeL{#1\hskip.31in}\pctsign}\global\advance\refno by1\findarg}
\def\findarg#1#{\begingroup\obeylines\newlinechar=`\^^M\pass@rg}
{\obeylines\gdef\pass@rg#1{\writ@line\relax #1^^M\hbox{}^^M}%
\gdef\writ@line#1^^M{\expandafter\toks0\expandafter{\striprel@x #1}%
\edef\next{\the\toks0}\ifx\next\em@rk\let\next=\endgroup\else\ifx\next\empty%
\else\immediate\write\wfile{\the\toks0}\fi\let\next=\writ@line\fi\next\relax}}
\def\striprel@x#1{} \def\em@rk{\hbox{}}
\def\lref{\begingroup\obeylines\lr@f}
\def\lr@f#1#2{\DefWarn#1\gdef#1{\let#1=\UNd@FiNeD\ref#1{#2}}\endgroup\unskip}

\def\addref#1{\immediate\write\rfile{\noexpand\item{}#1}} 
\def\listrefs{\footatend\vfill\supereject\immediate\closeout\rfile\writestoppt
\baselineskip=\footskip\centerline{{\bf References}}\bigskip{\parindent=20pt%
\frenchspacing\escapechar=` \input \jobname.refs\vfill\eject}\nonfrenchspacing}
\def\startrefs#1{\immediate\openout\rfile=\jobname.refs\refno=#1}
\def\xref{\expandafter\xr@f}\def\xr@f[#1]{#1}
\def\refs#1{\count255=1[\r@fs #1{\hbox{}}]}
\def\r@fs#1{\ifx\UNd@FiNeD#1\message{reflabel \string#1 is undefined.}%
\nref#1{need to supply reference \string#1.}\fi%
\vphantom{\hphantom{#1}}{\let\hyperref=\relax\xdef\next{#1}}%
\ifx\next\em@rk\def\next{}%
\else\ifx\next#1\ifodd\count255\relax\xref#1\count255=0\fi%
\else#1\count255=1\fi\let\next=\r@fs\fi\next}
%

%
\newwrite\ffile\global\newcount\figno \global\figno=1
\def\fig{fig.~\hyperref{}{figure}{\the\figno}{\the\figno}\nfig}
\def\nfig#1{\DefWarn#1%
\xdef#1{fig.~\noexpand\hyperref{}{figure}{\the\figno}{\the\figno}}%
\writedef{#1\leftbracket fig.\noexpand~\xfig#1}%
\ifnum\figno=1\immediate\openout\ffile=\jobname.figs\fi\chardef\wfile=\ffile%
{\let\hyperref=\relax
\immediate\write\ffile{\noexpand\medskip\noexpand\item{Fig.\ %
\noexpand\hyperdef\noexpand\hypernoname{figure}{\the\figno}{\the\figno}. }
\reflabeL{#1\hskip.55in}\pctsign}}\global\advance\figno by1\findarg}
\def\listfigs{\vfill\eject\immediate\closeout\ffile{\parindent40pt
\baselineskip14pt\centerline{{\bf Figure Captions}}\nobreak\medskip
\escapechar=` \input \jobname.figs\vfill\eject}}
\def\xfig{\expandafter\xf@g}\def\xf@g fig.\penalty\@M\ {}
\def\figs#1{figs.~\f@gs #1{\hbox{}}}
\def\f@gs#1{{\let\hyperref=\relax\xdef\next{#1}}\ifx\next\em@rk\def\next{}\else
\ifx\next#1\xfig #1\else#1\fi\let\next=\f@gs\fi\next}
\def\figin{\epsfcheck\figin}\def\figins{\epsfcheck\figins}
\def\epsfcheck{\ifx\epsfbox\UNd@FiNeD
\message{(NO epsf.tex, FIGURES WILL BE IGNORED)}
\gdef\figin##1{\vskip2in}\gdef\figins##1{\hskip.5in}
\else\message{(FIGURES WILL BE INCLUDED)}%
\gdef\figin##1{##1}\gdef\figins##1{##1}\fi}
\def\DefWarn#1{}
\def\figinsert{\goodbreak\midinsert}
\def\ifig#1#2#3{\DefWarn#1\xdef#1{fig.~\noexpand\hyperref{}{figure}%
{\the\figno}{\the\figno}}\writedef{#1\leftbracket fig.\noexpand~\xfig#1}%
\figinsert\figin{\centerline{#3}}\medskip\centerline{\vbox{\baselineskip12pt
\advance\hsize by -1truein\noindent\wrlabeL{#1=#1}\footnotefont%
{\bf Fig.~\hyperdef\hypernoname{figure}{\the\figno}{\the\figno}:} #2}}
\bigskip\endinsert\global\advance\figno by1}
\newwrite\lfile
{\escapechar-1\xdef\pctsign{\string\%}\xdef\leftbracket{\string\{}
\xdef\rightbracket{\string\}}\xdef\numbersign{\string\#}}
\def\writedefs{\immediate\openout\lfile=\jobname.defs \def\writedef##1{%
{\let\hyperref=\relax\let\hyperdef=\relax\let\hypernoname=\relax
 \immediate\write\lfile{\string\def\string##1\rightbracket}}}}%
\def\writestop{\def\writestoppt{\immediate\write\lfile{\string\pageno
 \the\pageno\string\startrefs\leftbracket\the\refno\rightbracket
 \string\def\string\secsym\leftbracket\secsym\rightbracket
 \string\secno\the\secno\string\meqno\the\meqno}\immediate\closeout\lfile}}
\def\writestoppt{}\def\writedef#1{}
\def\seclab#1{\DefWarn#1%
\xdef #1{\noexpand\hyperref{}{section}{\the\secno}{\the\secno}}%
\writedef{#1\leftbracket#1}\wrlabeL{#1=#1}}
\def\subseclab#1{\DefWarn#1%
\xdef #1{\noexpand\hyperref{}{subsection}{\secn@m.\the\subsecno}%
{\secn@m.\the\subsecno}}\writedef{#1\leftbracket#1}\wrlabeL{#1=#1}}
\def\applab#1{\DefWarn#1%
\xdef #1{\noexpand\hyperref{}{appendix}{\secn@m}{\secn@m}}%
\writedef{#1\leftbracket#1}\wrlabeL{#1=#1}}
\newwrite\tfile \def\writetoca#1{}
\def\leaderfill{\leaders\hbox to 1em{\hss.\hss}\hfill}
\def\writetoc{\immediate\openout\tfile=\jobname.toc
   \def\writetoca##1{{\edef\next{\write\tfile{\noindent ##1
   \string\leaderfill {\string\hyperref{}{page}{\noexpand\number\pageno}%
                       {\noexpand\number\pageno}} \par}}\next}}}
\newread\ch@ckfile
\def\listtoc{\immediate\closeout\tfile\immediate\openin\ch@ckfile=\jobname.toc
\ifeof\ch@ckfile\message{no file \jobname.toc, no table of contents this pass}%
\else\closein\ch@ckfile\centerline{\bf Contents}\nobreak\medskip%
{\baselineskip=12pt\footnotefont\parskip=0pt\catcode`\@=11\input\jobname.toc
\catcode`\@=12\bigbreak\bigskip}\fi}
\catcode`\@=12 
%
\edef\tfontsize{\ifx\answ\bigans scaled\magstep3\else scaled\magstep4\fi}
\font\titlerm=cmr10 \tfontsize \font\titlerms=cmr7 \tfontsize
\font\titlermss=cmr5 \tfontsize \font\titlei=cmmi10 \tfontsize
\font\titleis=cmmi7 \tfontsize \font\titleiss=cmmi5 \tfontsize
\font\titlesy=cmsy10 \tfontsize \font\titlesys=cmsy7 \tfontsize
\font\titlesyss=cmsy5 \tfontsize \font\titleit=cmti10 \tfontsize
\skewchar\titlei='177 \skewchar\titleis='177 \skewchar\titleiss='177
\skewchar\titlesy='60 \skewchar\titlesys='60 \skewchar\titlesyss='60
\def\titlefont{\def\rm{\fam0\titlerm}
\textfont0=\titlerm \scriptfont0=\titlerms \scriptscriptfont0=\titlermss
\textfont1=\titlei \scriptfont1=\titleis \scriptscriptfont1=\titleiss
\textfont2=\titlesy \scriptfont2=\titlesys \scriptscriptfont2=\titlesyss
\textfont\itfam=\titleit \def\it{\fam\itfam\titleit}\rm}
 \ifx\answ\bigans\else scaled\magstep1\fi
\ifx\answ\bigans\def\abstractfont{\tenpoint}\else
\font\absit=cmti10 scaled \magstep1
\font\abssl=cmsl10 scaled \magstep1
\font\absrm=cmr10 scaled\magstep1 \font\absrms=cmr7 scaled\magstep1
\font\absrmss=cmr5 scaled\magstep1 \font\absi=cmmi10 scaled\magstep1
\font\absis=cmmi7 scaled\magstep1 \font\absiss=cmmi5 scaled\magstep1
\font\abssy=cmsy10 scaled\magstep1 \font\abssys=cmsy7 scaled\magstep1
\font\abssyss=cmsy5 scaled\magstep1 \font\absbf=cmbx10 scaled\magstep1
\skewchar\absi='177 \skewchar\absis='177 \skewchar\absiss='177
\skewchar\abssy='60 \skewchar\abssys='60 \skewchar\abssyss='60
\def\abstractfont{\def\rm{\fam0\absrm}
\textfont0=\absrm \scriptfont0=\absrms \scriptscriptfont0=\absrmss
\textfont1=\absi \scriptfont1=\absis \scriptscriptfont1=\absiss
\textfont2=\abssy \scriptfont2=\abssys \scriptscriptfont2=\abssyss
\textfont\itfam=\absit \def\it{\fam\itfam\absit}\def\footnotefont{\tenpoint}%
\textfont\slfam=\abssl \def\sl{\fam\slfam\abssl}%
\textfont\bffam=\absbf \def\bf
{\fam\bffam\absbf}\rm}\fi
\def\tenpoint{\def\rm{\fam0\tenrm}
\textfont0=\tenrm \scriptfont0=\sevenrm \scriptscriptfont0=\fiverm
\textfont1=\teni  \scriptfont1=\seveni  \scriptscriptfont1=\fivei
\textfont2=\tensy \scriptfont2=\sevensy \scriptscriptfont2=\fivesy
\textfont\itfam=\tenit \def\it{\fam\itfam\tenit}\def\footnotefont{\ninepoint}%
\textfont\bffam=\tenbf \def\bf{\fam\bffam\tenbf}\def\sl{\fam\slfam\tensl}\rm}
\font\ninerm=cmr9 \font\sixrm=cmr6 \font\ninei=cmmi9 \font\sixi=cmmi6
\font\ninesy=cmsy9 \font\sixsy=cmsy6 \font\ninebf=cmbx9
\font\nineit=cmti9 \font\ninesl=cmsl9 \skewchar\ninei='177
\skewchar\sixi='177 \skewchar\ninesy='60 \skewchar\sixsy='60
\def\ninepoint{\def\rm{\fam0\ninerm}
\textfont0=\ninerm \scriptfont0=\sixrm \scriptscriptfont0=\fiverm
\textfont1=\ninei \scriptfont1=\sixi \scriptscriptfont1=\fivei
\textfont2=\ninesy \scriptfont2=\sixsy \scriptscriptfont2=\fivesy
\textfont\itfam=\ninei \def\it{\fam\itfam\nineit}\def\sl{\fam\slfam\ninesl}%
\textfont\bffam=\ninebf \def\bf{\fam\bffam\ninebf}\rm}
%
%
\def\noblackbox{\overfullrule=0pt}
\hyphenation{anom-aly anom-alies coun-ter-term coun-ter-terms}
\def\inv{^{\raise.15ex\hbox{${\scriptscriptstyle -}$}\kern-.05em 1}}

\def\Dsl{\,\raise.15ex\hbox{/}\mkern-13.5mu D} 
\def\dsl{\raise.15ex\hbox{/}\kern-.57em\partial}
\def\del{\partial}

\def\lspace{\ifx\answ\bigans{}\else\qquad\fi}
\def\lbspace{\ifx\answ\bigans{}\else\hskip-.2in\fi} 

\def\boxeqn#1{\vcenter{\vbox{\hrule\hbox{\vrule\kern3pt\vbox{\kern3pt
	\hbox{${\displaystyle #1}$}\kern3pt}\kern3pt\vrule}\hrule}}}
\def\mbox#1#2{\vcenter{\hrule \hbox{\vrule height#2in
		\kern#1in \vrule} \hrule}}  

\def\darr#1{\raise1.5ex\hbox{$\leftrightarrow$}\mkern-16.5mu #1}

\def\roughly#1{\raise.3ex\hbox{$#1$\kern-.75em\lower1ex\hbox{$\sim$}}}

\input amssym
\input epsf

\def\IZ{\relax\ifmmode\mathchoice
{\hbox{\cmss Z\kern-.4em Z}}{\hbox{\cmss Z\kern-.4em Z}} {\lower.9pt\hbox{\cmsss Z\kern-.4em Z}}
{\lower1.2pt\hbox{\cmsss Z\kern-.4em Z}}\else{\cmss Z\kern-.4em Z}\fi}

\newif\ifdraft\draftfalse
\newif\ifinter\interfalse
\ifdraft\draftmode\else\interfalse\fi
\def\journal#1&#2(#3){\unskip, \sl #1\ \bf #2 \rm(19#3) }
\def\andjournal#1&#2(#3){\sl #1~\bf #2 \rm (19#3) }

\def\frac#1#2{{#1\over#2}}

\def\ds{\displaystyle}

\def\inbar{\,\vrule height1.5ex width.4pt depth0pt}
\def\IC{\relax\hbox{$\inbar\kern-.3em{\rm C}$}}
\def\IR{\relax{\rm I\kern-.18em R}}
\def\IP{\relax{\rm I\kern-.18em P}}

%
%


%
\catcode`\@=11
\def\slash#1{\mathord{\mathpalette\c@ncel{#1}}}
\overfullrule=0pt

\def\Z{\hbox{$\bb Z$}}
\def\R{\hbox{$\bb R$}}

\def\underrel#1\over#2{\mathrel{\mathop{\kern\z@#1}\limits_{#2}}}

\catcode`\@=12


%


\def\[{[}
\def\]{]}

\def\comment#1{ }

%
\def\draftnote#1{\ifdraft{\baselineskip2ex
                 \vbox{\kern1em\hrule\hbox{\vrule\kern1em\vbox{\kern1ex
                 \noindent \underbar{NOTE}: #1
             \vskip1ex}\kern1em\vrule}\hrule}}\fi}
\def\internote#1{\ifinter{\baselineskip2ex
                 \vbox{\kern1em\hrule\hbox{\vrule\kern1em\vbox{\kern1ex
                 \noindent \underbar{Internal Note}: #1
             \vskip1ex}\kern1em\vrule}\hrule}}\fi}

%

%
%

%

\def\inv{^{-1}}



\def\b{\beta}


\def\bb{
\font\tenmsb=msbm10
\font\sevenmsb=msbm7
\font\fivemsb=msbm5
\textfont1=\tenmsb
\scriptfont1=\sevenmsb
\scriptscriptfont1=\fivemsb
}





\def\bar{\overline}
\def\b{\bar}
\def\bsq#1{{{\b{#1}}^{\lower 2.5pt\hbox{$\scriptstyle 2$}}}}
\def\bexp#1#2{{{\b{#1}}^{\lower 2.5pt\hbox{$\scriptstyle #2$}}}}
\def\dotexp#1#2{{{#1}^{\lower 2.5pt\hbox{$\scriptstyle #2$}}}}


\def\sign{\mathop{\rm sgn}}

\def\rt2{\sqrt{2}}



\def\CM{{\cal M}}
\def\CN{{\cal N}}
\def\CO{{\cal O}}

\def\CV{{\cal V}}


\def\1{{\ds 1}}
\def\R{\hbox{$\bb R$}}
\def\C{\hbox{$\bb C$}}

\def\Z{\hbox{$\bb Z$}}

\def\P{\hbox{$\bb P$}}


\noblackbox

\def\unit{\relax{\rm 1\kern-.26em I}}
\def\nada{\relax{\rm 0\kern-.30em l}}

\noblackbox
\def\IL{\relax{\rm I\kern-.18em L}}
\def\IH{\relax{\rm I\kern-.18em H}}
\def\IR{\relax{\rm I\kern-.18em R}}
\def\IC{\relax\hbox{$\inbar\kern-.3em{\rm C}$}}
\def\IZ{\relax\ifmmode\mathchoice
{\hbox{\cmss Z\kern-.4em Z}}{\hbox{\cmss Z\kern-.4em Z}} {\lower.9pt\hbox{\cmsss Z\kern-.4em Z}}
{\lower1.2pt\hbox{\cmsss Z\kern-.4em Z}}\else{\cmss Z\kern-.4em Z}\fi}
\def\CM {{\cal M}}

\def\partialslash{\not{\hbox{\kern-2pt $\partial$}}}

\font\manual=manfnt \def\dbend{\lower3.5pt\hbox{\manual\char127}}

\def\IZ{\relax\ifmmode\mathchoice
{\hbox{\cmss Z\kern-.4em Z}}{\hbox{\cmss Z\kern-.4em Z}} {\lower.9pt\hbox{\cmsss Z\kern-.4em Z}}
{\lower1.2pt\hbox{\cmsss Z\kern-.4em Z}}\else{\cmss Z\kern-.4em Z}\fi}
\def\half{{1\over 2}}

\def\bar{\overline}

\def\rt2{\sqrt{2}}
\def\irt2{{1\over\sqrt{2}}}

\def\slashchar#1{\setbox0=\hbox{$#1$}           
   \dimen0=\wd0                                 
   \setbox1=\hbox{/} \dimen1=\wd1               
   \ifdim\dimen0>\dimen1                        
      \rlap{\hbox to \dimen0{\hfil/\hfil}}      
      #1                                        
   \else                                        
      \rlap{\hbox to \dimen1{\hfil$#1$\hfil}}   
      /                                         
   \fi}


\def\figcaption#1#2{\DefWarn#1\xdef#1{Figure~\noexpand\hyperref{}{figure}%
{\the\figno}{\the\figno}}\writedef{#1\leftbracket Figure\noexpand~\xfig#1}%
\medskip\centerline{{\footnotefont\bf Figure~\hyperdef\hypernoname{figure}{\the\figno}{\the\figno}:}  #2 \wrlabeL{#1=#1}}%
\global\advance\figno by1}

%
%

\lref\AbanovQZ{
  A.~G.~Abanov and P.~B.~Wiegmann,
  ``Theta terms in nonlinear sigma models,''
Nucl.\ Phys.\ B {\bf 570}, 685 (2000).
[hep-th/9911025].
}

\lref\AffleckAS{
  I.~Affleck, J.~A.~Harvey and E.~Witten,
  ``Instantons and (Super)Symmetry Breaking in (2+1)-Dimensions,''
Nucl.\ Phys.\ B {\bf 206}, 413 (1982).
}

\lref\AharonyBX{
  O.~Aharony, A.~Hanany, K.~A.~Intriligator, N.~Seiberg and M.~J.~Strassler,
  ``Aspects of $N=2$ supersymmetric gauge theories in three-dimensions,''
Nucl.\ Phys.\ B {\bf 499}, 67 (1997).
[hep-th/9703110].
}

\lref\AharonyGP{
  O.~Aharony,
  ``IR duality in $d = 3$ $N=2$ supersymmetric $USp(2N_c)$ and $U(N_c)$ gauge theories,''
Phys.\ Lett.\ B {\bf 404}, 71 (1997).
[hep-th/9703215].
}

\lref\AharonyCI{
  O.~Aharony and I.~Shamir,
  ``On $O(N_c)$ $d=3$ ${\cal N}{=}2$ supersymmetric QCD Theories,''
JHEP {\bf 1112}, 043 (2011).
[arXiv:1109.5081 [hep-th]].
}

\lref\AharonyJZ{
  O.~Aharony, G.~Gur-Ari and R.~Yacoby,
  ``$d=3$ Bosonic Vector Models Coupled to Chern-Simons Gauge Theories,''
JHEP {\bf 1203}, 037 (2012).
[arXiv:1110.4382 [hep-th]].
}

\lref\AharonyNH{
  O.~Aharony, G.~Gur-Ari and R.~Yacoby,
  ``Correlation Functions of Large $N$ Chern-Simons-Matter Theories and Bosonization in Three Dimensions,''
JHEP {\bf 1212}, 028 (2012).
[arXiv:1207.4593 [hep-th]].
}

\lref\AharonyNS{
  O.~Aharony, S.~Giombi, G.~Gur-Ari, J.~Maldacena and R.~Yacoby,
  ``The Thermal Free Energy in Large $N$ Chern-Simons-Matter Theories,''
JHEP {\bf 1303}, 121 (2013).
[arXiv:1211.4843 [hep-th]].
}

\lref\AharonyHDA{
  O.~Aharony, N.~Seiberg and Y.~Tachikawa,
  ``Reading between the lines of four-dimensional gauge theories,''
JHEP {\bf 1308}, 115 (2013).
[arXiv:1305.0318 [hep-th]].
}

\lref\AharonyDHA{
  O.~Aharony, S.~S.~Razamat, N.~Seiberg and B.~Willett,
  ``3d dualities from 4d dualities,''
JHEP {\bf 1307}, 149 (2013).
[arXiv:1305.3924 [hep-th]].
}

\lref\AharonyKMA{
  O.~Aharony, S.~S.~Razamat, N.~Seiberg and B.~Willett,
  ``3d dualities from 4d dualities for orthogonal groups,''
JHEP {\bf 1308}, 099 (2013)
[arXiv:1307.0511 [hep-th]].
}

\lref\AharonyMJS{
  O.~Aharony,
  ``Baryons, monopoles and dualities in Chern-Simons-matter theories,''
JHEP {\bf 1602}, 093 (2016).
[arXiv:1512.00161 [hep-th]].
}

\lref\AharonyJVV{
  O.~Aharony, F.~Benini, P.~S.~Hsin and N.~Seiberg,
  ``Chern-Simons-matter dualities with $SO$ and $USp$ gauge groups,''
JHEP {\bf 1702}, 072 (2017).
[arXiv:1611.07874 [cond-mat.str-el]].
}

\lref\WangTXT{
  C.~Wang, A.~Nahum, M.~A.~Metlitski, C.~Xu and T.~Senthil,
  ``Deconfined quantum critical points: symmetries and dualities,''
[arXiv:1703.02426 [cond-mat.str-el]].
}

\lref\AlvarezGaumeNF{
  L.~Alvarez-Gaume, S.~Della Pietra and G.~W.~Moore,
  ``Anomalies and Odd Dimensions,''
Annals Phys.\  {\bf 163}, 288 (1985).
}

\lref\AnninosUI{
  D.~Anninos, T.~Hartman and A.~Strominger,
  ``Higher Spin Realization of the dS/CFT Correspondence,''
[arXiv:1108.5735 [hep-th]].
}

\lref\AnninosHIA{
  D.~Anninos, R.~Mahajan, D�.~Radicevic and E.~Shaghoulian,
  ``Chern-Simons-Ghost Theories and de Sitter Space,''
JHEP {\bf 1501}, 074 (2015).
[arXiv:1405.1424 [hep-th]].
}

\lref\AtiyahJF{
  M.~F.~Atiyah, V.~K.~Patodi and I.~M.~Singer,
  ``Spectral Asymmetry in Riemannian Geometry, I,''
  Math.\ Proc.\ Camb.\ Phil.\ Soc.\ {\bf 77} (1975) 43--69.
}

\lref\BanksZN{
  T.~Banks and N.~Seiberg,
  ``Symmetries and Strings in Field Theory and Gravity,''
Phys.\ Rev.\ D {\bf 83}, 084019 (2011).
[arXiv:1011.5120 [hep-th]].
}

\lref\BarkeshliIDA{
  M.~Barkeshli and J.~McGreevy,
  ``Continuous transition between fractional quantum Hall and superfluid states,''
Phys.\ Rev.\ B {\bf 89}, 235116 (2014).
}

\lref\BeemMB{
  C.~Beem, T.~Dimofte and S.~Pasquetti,
  ``Holomorphic Blocks in Three Dimensions,''
[arXiv:1211.1986 [hep-th]].
}
\lref\VafaXG{
  C.~Vafa and E.~Witten,
  ``Parity Conservation in QCD,''
Phys.\ Rev.\ Lett.\  {\bf 53}, 535 (1984).
}

\lref\BeniniMF{
  F.~Benini, C.~Closset and S.~Cremonesi,
  ``Comments on 3d Seiberg-like dualities,''
JHEP {\bf 1110}, 075 (2011).
[arXiv:1108.5373 [hep-th]].
}

\lref\BernardXY{
  D.~Bernard,
  ``String Characters From {Kac-Moody} Automorphisms,''
  Nucl.\ Phys.\ B {\bf 288}, 628 (1987).
}

\lref\BhattacharyaZY{
  J.~Bhattacharya, S.~Bhattacharyya, S.~Minwalla and S.~Raju,
  ``Indices for Superconformal Field Theories in 3,5 and 6 Dimensions,''
JHEP {\bf 0802}, 064 (2008).
[arXiv:0801.1435 [hep-th]].
}

\lref\deBoerMP{
  J.~de Boer, K.~Hori, H.~Ooguri and Y.~Oz,
  ``Mirror symmetry in three-dimensional gauge theories, quivers and D-branes,''
Nucl.\ Phys.\ B {\bf 493}, 101 (1997).
[hep-th/9611063].
}

\lref\deBoerKA{
  J.~de Boer, K.~Hori, Y.~Oz and Z.~Yin,
  ``Branes and mirror symmetry in N=2 supersymmetric gauge theories in three-dimensions,''
Nucl.\ Phys.\ B {\bf 502}, 107 (1997).
[hep-th/9702154].
}

\lref\WilczekCY{
  F.~Wilczek and A.~Zee,
  ``Linking Numbers, Spin, and Statistics of Solitons,''
Phys.\ Rev.\ Lett.\  {\bf 51}, 2250 (1983).
}

\lref\BondersonPLA{
  P.~Bonderson, C.~Nayak and X.~L.~Qi,
  ``A time-reversal invariant topological phase at the surface of a 3D topological insulator,''
J.\ Stat.\ Mech.\  {\bf 2013}, P09016 (2013).
}

\lref\BorokhovIB{
  V.~Borokhov, A.~Kapustin and X.~k.~Wu,
  ``Topological disorder operators in three-dimensional conformal field theory,''
JHEP {\bf 0211}, 049 (2002).
[hep-th/0206054].
}

\lref\WittenTW{
  E.~Witten,
  ``Global Aspects of Current Algebra,''
Nucl.\ Phys.\ B {\bf 223}, 422 (1983).
}
\lref\WittenTX{
  E.~Witten,
  ``Current Algebra, Baryons, and Quark Confinement,''
Nucl.\ Phys.\ B {\bf 223}, 433 (1983).
}

\lref\GaiottoYUP{
  D.~Gaiotto, A.~Kapustin, Z.~Komargodski and N.~Seiberg,
  ``Theta, Time Reversal, and Temperature,''
[arXiv:1703.00501 [hep-th]].
}

\lref\BorokhovCG{
  V.~Borokhov, A.~Kapustin and X.~k.~Wu,
  ``Monopole operators and mirror symmetry in three-dimensions,''
JHEP {\bf 0212}, 044 (2002).
[hep-th/0207074].
}

\lref\KomargodskiDMC{
  Z.~Komargodski, A.~Sharon, R.~Thorngren and X.~Zhou,
  ``Comments on Abelian Higgs Models and Persistent Order,''
[arXiv:1705.04786 [hep-th]].
}

\lref\KomargodskiSMK{
  Z.~Komargodski, T.~Sulejmanpasic and M.~Unsal,
  ``Walls, Anomalies, and (De) Confinement in Quantum Anti-Ferromagnets,''
[arXiv:1706.05731 [cond-mat.str-el]].
}

\lref\Browder{
  W.~Browder and E.~Thomas,
  ``Axioms for the generalized Pontryagin cohomology operations,''
  Quart.\ J.\ Math.\ Oxford {\bf 13}, 55--60 (1962).
}

\lref\debult{
  F.~van~de~Bult,
  ``Hyperbolic Hypergeometric Functions,''
University of Amsterdam Ph.D. thesis
}

\lref\Camperi{
  M.~Camperi, F.~Levstein and G.~Zemba,
  ``The Large N Limit Of Chern-simons Gauge Theory,''
  Phys.\ Lett.\ B {\bf 247} (1990) 549.
}

\lref\ChenCD{
  W.~Chen, M.~P.~A.~Fisher and Y.~S.~Wu,
  ``Mott transition in an anyon gas,''
Phys.\ Rev.\ B {\bf 48}, 13749 (1993).
[cond-mat/9301037].
}

\lref\WittenDS{
  E.~Witten,
  ``Supersymmetric index of three-dimensional gauge theory,''
In *Shifman, M.A. (ed.): The many faces of the superworld* 156-184.
[hep-th/9903005].
}

\lref\ChenJHA{
  X.~Chen, L.~Fidkowski and A.~Vishwanath,
  ``Symmetry Enforced Non-Abelian Topological Order at the Surface of a Topological Insulator,''
Phys.\ Rev.\ B {\bf 89}, no. 16, 165132 (2014).
[arXiv:1306.3250 [cond-mat.str-el]].
}

\lref\IntriligatorLCA{
  K.~Intriligator and N.~Seiberg,
  ``Aspects of 3d N=2 Chern-Simons-Matter Theories,''
JHEP {\bf 1307}, 079 (2013).
[arXiv:1305.1633 [hep-th]].
}

\lref\ChengPDN{
  M.~Cheng and C.~Xu,
  ``Series of (2+1)-dimensional stable self-dual interacting conformal field theories,''
Phys.\ Rev.\ B {\bf 94}, 214415 (2016). 
[arXiv:1609.02560 [cond-mat.str-el]].
}

\lref\ClossetVG{
  C.~Closset, T.~T.~Dumitrescu, G.~Festuccia, Z.~Komargodski and N.~Seiberg,
  ``Contact Terms, Unitarity, and F-Maximization in Three-Dimensional Superconformal Theories,''
JHEP {\bf 1210}, 053 (2012).
[arXiv:1205.4142 [hep-th]].
}

\lref\ClossetVP{
  C.~Closset, T.~T.~Dumitrescu, G.~Festuccia, Z.~Komargodski and N.~Seiberg,
  ``Comments on Chern-Simons Contact Terms in Three Dimensions,''
JHEP {\bf 1209}, 091 (2012).
[arXiv:1206.5218 [hep-th]].
}

\lref\ClossetRU{
  C.~Closset, T.~T.~Dumitrescu, G.~Festuccia and Z.~Komargodski,
  ``Supersymmetric Field Theories on Three-Manifolds,''
JHEP {\bf 1305}, 017 (2013).
[arXiv:1212.3388 [hep-th]].
}

\lref\CveticXN{
  M.~Cvetic, T.~W.~Grimm and D.~Klevers,
  ``Anomaly Cancellation And Abelian Gauge Symmetries In F-theory,''
JHEP {\bf 1302}, 101 (2013).
[arXiv:1210.6034 [hep-th]].
}

\lref\DaiKQ{
  X.~z.~Dai and D.~S.~Freed,
  ``eta invariants and determinant lines,''
J.\ Math.\ Phys.\  {\bf 35}, 5155 (1994), Erratum: [J.\ Math.\ Phys.\  {\bf 42}, 2343 (2001)].
[hep-th/9405012].
}

\lref\DasguptaZZ{
  C.~Dasgupta and B.~I.~Halperin,
  ``Phase Transition in a Lattice Model of Superconductivity,''
Phys.\ Rev.\ Lett.\  {\bf 47}, 1556 (1981).
}

\lref\DaviesUW{
  N.~M.~Davies, T.~J.~Hollowood, V.~V.~Khoze and M.~P.~Mattis,
  ``Gluino condensate and magnetic monopoles in supersymmetric gluodynamics,''
Nucl.\ Phys.\ B {\bf 559}, 123 (1999).
[hep-th/9905015].
}

\lref\DaviesNW{
  N.~M.~Davies, T.~J.~Hollowood and V.~V.~Khoze,
  ``Monopoles, affine algebras and the gluino condensate,''
J.\ Math.\ Phys.\  {\bf 44}, 3640 (2003).
[hep-th/0006011].
}

\lref\DimoftePY{
  T.~Dimofte, D.~Gaiotto and S.~Gukov,
  ``3-Manifolds and 3d Indices,''
[arXiv:1112.5179 [hep-th]].
}

\lref\DolanQI{
  F.~A.~Dolan and H.~Osborn,
  ``Applications of the Superconformal Index for Protected Operators and q-Hypergeometric Identities to N=1 Dual Theories,''
Nucl.\ Phys.\ B {\bf 818}, 137 (2009).
[arXiv:0801.4947 [hep-th]].
}

\lref\DolanRP{
  F.~A.~H.~Dolan, V.~P.~Spiridonov and G.~S.~Vartanov,
  ``From 4d superconformal indices to 3d partition functions,''
Phys.\ Lett.\ B {\bf 704}, 234 (2011).
[arXiv:1104.1787 [hep-th]].
}

\lref\DouglasEX{
  M.~R.~Douglas,
  ``Chern-Simons-Witten theory as a topological Fermi liquid,''
[hep-th/9403119].
}

\lref\EagerHX{
  R.~Eager, J.~Schmude and Y.~Tachikawa,
  ``Superconformal Indices, Sasaki-Einstein Manifolds, and Cyclic Homologies,''
[arXiv:1207.0573 [hep-th]].
}

\lref\ElitzurFH{
  S.~Elitzur, A.~Giveon and D.~Kutasov,
  ``Branes and N=1 duality in string theory,''
Phys.\ Lett.\ B {\bf 400}, 269 (1997).
[hep-th/9702014].
}

\lref\KapustinLWA{
  A.~Kapustin and R.~Thorngren,
  ``Anomalies of discrete symmetries in three dimensions and group cohomology,''
Phys.\ Rev.\ Lett.\  {\bf 112}, no. 23, 231602 (2014).
[arXiv:1403.0617 [hep-th]].
}
\lref\KapustinZVA{
  A.~Kapustin and R.~Thorngren,
  ``Anomalies of discrete symmetries in various dimensions and group cohomology,''
[arXiv:1404.3230 [hep-th]].
}

\lref\ElitzurHC{
  S.~Elitzur, A.~Giveon, D.~Kutasov, E.~Rabinovici and A.~Schwimmer,
  ``Brane dynamics and N=1 supersymmetric gauge theory,''
Nucl.\ Phys.\ B {\bf 505}, 202 (1997).
[hep-th/9704104].
}

\lref\EssinRQ{
  A.~M.~Essin, J.~E.~Moore and D.~Vanderbilt,
  ``Magnetoelectric polarizability and axion electrodynamics in crystalline insulators,''
Phys.\ Rev.\ Lett.\  {\bf 102}, 146805 (2009).
[arXiv:0810.2998 [cond-mat.mes-hall]].
}

\lref\slthreeZ{
  J.~Felder, A.~Varchenko,
  ``The elliptic gamma function and $SL(3,Z) \times Z^3$,'' $\;\;$
[arXiv:math/0001184].
}

\lref\FestucciaWS{
  G.~Festuccia and N.~Seiberg,
  ``Rigid Supersymmetric Theories in Curved Superspace,''
JHEP {\bf 1106}, 114 (2011).
[arXiv:1105.0689 [hep-th]].
}

\lref\FidkowskiJUA{
  L.~Fidkowski, X.~Chen and A.~Vishwanath,
  ``Non-Abelian Topological Order on the Surface of a 3D Topological Superconductor from an Exactly Solved Model,''
Phys.\ Rev.\ X {\bf 3}, no. 4, 041016 (2013).
[arXiv:1305.5851 [cond-mat.str-el]].
}

\lref\FradkinTT{
  E.~H.~Fradkin and F.~A.~Schaposnik,
  ``The Fermion - boson mapping in three-dimensional quantum field theory,''
Phys.\ Lett.\ B {\bf 338}, 253 (1994).
[hep-th/9407182].
}

\lref\GaddeEN{
  A.~Gadde, L.~Rastelli, S.~S.~Razamat and W.~Yan,
  ``On the Superconformal Index of N=1 IR Fixed Points: A Holographic Check,''
JHEP {\bf 1103}, 041 (2011).
[arXiv:1011.5278 [hep-th]].
}

\lref\HongSB{
  D.~K.~Hong and H.~U.~Yee,
  ``Holographic aspects of three dimensional QCD from string theory,''
JHEP {\bf 1005}, 036 (2010), Erratum: [JHEP {\bf 1008}, 120 (2010)].
[arXiv:1003.1306 [hep-th]].
}

\lref\GaddeIA{
  A.~Gadde and W.~Yan,
  ``Reducing the 4d Index to the $S^3$ Partition Function,''
JHEP {\bf 1212}, 003 (2012).
[arXiv:1104.2592 [hep-th]].
}

\lref\GaddeDDA{
  A.~Gadde and S.~Gukov,
  ``2d Index and Surface operators,''
[arXiv:1305.0266 [hep-th]].
}

\lref\VafaXH{
  C.~Vafa and E.~Witten,
  ``Eigenvalue Inequalities for Fermions in Gauge Theories,''
Commun.\ Math.\ Phys.\  {\bf 95}, 257 (1984).
}

\lref\GaiottoAK{
  D.~Gaiotto and E.~Witten,
  ``S-Duality of Boundary Conditions In N=4 Super Yang-Mills Theory,''
Adv.\ Theor.\ Math.\ Phys.\  {\bf 13}, no. 3, 721 (2009).
[arXiv:0807.3720 [hep-th]].
}

\lref\GaiottoBE{
  D.~Gaiotto, G.~W.~Moore and A.~Neitzke,
  ``Framed BPS States,''
[arXiv:1006.0146 [hep-th]].
}

\lref\GaiottoKFA{
  D.~Gaiotto, A.~Kapustin, N.~Seiberg and B.~Willett,
  ``Generalized Global Symmetries,''
JHEP {\bf 1502}, 172 (2015).
[arXiv:1412.5148 [hep-th]].
}

\lref\GeraedtsPVA{
  S.~D.~Geraedts, M.~P.~Zaletel, R.~S.~K.~Mong, M.~A.~Metlitski, A.~Vishwanath and O.~I.~Motrunich,
  ``The half-filled Landau level: the case for Dirac composite fermions,''
Science {\bf 352}, 197 (2016).
[arXiv:1508.04140 [cond-mat.str-el]].
}

\lref\GiombiYA{
  S.~Giombi and X.~Yin,
  ``On Higher Spin Gauge Theory and the Critical $O(N)$ Model,''
Phys.\ Rev.\ D {\bf 85}, 086005 (2012).
[arXiv:1105.4011 [hep-th]].
}

\lref\GiombiKC{
  S.~Giombi, S.~Minwalla, S.~Prakash, S.~P.~Trivedi, S.~R.~Wadia and X.~Yin,
  ``Chern-Simons Theory with Vector Fermion Matter,''
Eur.\ Phys.\ J.\ C {\bf 72}, 2112 (2012).
[arXiv:1110.4386 [hep-th]].
}

\lref\GiombiMS{
  S.~Giombi and X.~Yin,
  ``The Higher Spin/Vector Model Duality,''
J.\ Phys.\ A {\bf 46}, 214003 (2013).
[arXiv:1208.4036 [hep-th]].
}

\lref\GiombiZWA{
  S.~Giombi, V.~Gurucharan, V.~Kirilin, S.~Prakash and E.~Skvortsov,
  ``On the Higher-Spin Spectrum in Large $N$ Chern-Simons Vector Models,''
JHEP {\bf 1701}, 058 (2017).
[arXiv:1610.08472 [hep-th]].
}

\lref\GiveonSR{
  A.~Giveon and D.~Kutasov,
  ``Brane dynamics and gauge theory,''
Rev.\ Mod.\ Phys.\  {\bf 71}, 983 (1999).
[hep-th/9802067].
}

\lref\GiveonZN{
  A.~Giveon and D.~Kutasov,
  ``Seiberg Duality in Chern-Simons Theory,''
Nucl.\ Phys.\ B {\bf 812}, 1 (2009).
[arXiv:0808.0360 [hep-th]].
}

\lref\GoddardQE{
  P.~Goddard, J.~Nuyts and D.~I.~Olive,
  ``Gauge Theories and Magnetic Charge,''
Nucl.\ Phys.\ B {\bf 125}, 1 (1977).
}

\lref\GoddardVK{
  P.~Goddard, A.~Kent and D.~I.~Olive,
  ``Virasoro Algebras and Coset Space Models,''
Phys.\ Lett.\ B {\bf 152}, 88 (1985).
}

\lref\GreenDA{
  D.~Green, Z.~Komargodski, N.~Seiberg, Y.~Tachikawa and B.~Wecht,
  ``Exactly Marginal Deformations and Global Symmetries,''
JHEP {\bf 1006}, 106 (2010).
[arXiv:1005.3546 [hep-th]].
}

\lref\GurPCA{
  G.~Gur-Ari and R.~Yacoby,
  ``Three Dimensional Bosonization From Supersymmetry,''
JHEP {\bf 1511}, 013 (2015).
[arXiv:1507.04378 [hep-th]].
}

\lref\GurAriXFF{
  G.~Gur-Ari, S.~A.~Hartnoll and R.~Mahajan,
  ``Transport in Chern-Simons-Matter Theories,''
JHEP {\bf 1607}, 090 (2016).
[arXiv:1605.01122 [hep-th]].
}

\lref\HalperinMH{
  B.~I.~Halperin, P.~A.~Lee and N.~Read,
  ``Theory of the half filled Landau level,''
Phys.\ Rev.\ B {\bf 47}, 7312 (1993).
}

\lref\HamaEA{
  N.~Hama, K.~Hosomichi and S.~Lee,
  ``SUSY Gauge Theories on Squashed Three-Spheres,''
JHEP {\bf 1105}, 014 (2011).
[arXiv:1102.4716 [hep-th]].
}

\lref\Hasegawa{
K.~Hasegawa,
  ``Spin Module Versions of Weyl's Reciprocity Theorem for Classical Kac-Moody Lie Algebras - An Application to Branching Rule Duality,''
Publ.\ Res.\ Inst.\ Math.\ Sci.\ {\bf 25}, 741-828 (1989).
}

\lref\HoriDK{
  K.~Hori and D.~Tong,
  ``Aspects of Non-Abelian Gauge Dynamics in Two-Dimensional N=(2,2) Theories,''
JHEP {\bf 0705}, 079 (2007).
[hep-th/0609032].
}

\lref\HoriPD{
  K.~Hori,
  ``Duality In Two-Dimensional (2,2) Supersymmetric Non-Abelian Gauge Theories,''
[arXiv:1104.2853 [hep-th]].
}

\lref\HsinBLU{
  P.~S.~Hsin and N.~Seiberg,
  ``Level/rank Duality and Chern-Simons-Matter Theories,''
JHEP {\bf 1609}, 095 (2016).
[arXiv:1607.07457 [hep-th]].
}

\lref\HullMS{
  C.~M.~Hull and B.~J.~Spence,
  ``The Geometry of the gauged sigma model with Wess-Zumino term,''
Nucl.\ Phys.\ B {\bf 353}, 379 (1991).
}

\lref\HwangQT{
  C.~Hwang, H.~Kim, K.~-J.~Park and J.~Park,
  ``Index computation for 3d Chern-Simons matter theory: test of Seiberg-like duality,''
JHEP {\bf 1109}, 037 (2011).
[arXiv:1107.4942 [hep-th]].
}

\lref\HwangHT{
  C.~Hwang, K.~-J.~Park and J.~Park,
  ``Evidence for Aharony duality for orthogonal gauge groups,''
JHEP {\bf 1111}, 011 (2011).
[arXiv:1109.2828 [hep-th]].
}

\lref\HwangJH{
  C.~Hwang, H.~-C.~Kim and J.~Park,
  ``Factorization of the 3d superconformal index,''
[arXiv:1211.6023 [hep-th]].
}

\lref\ImamuraSU{
  Y.~Imamura and S.~Yokoyama,
  ``Index for three dimensional superconformal field theories with general R-charge assignments,''
JHEP {\bf 1104}, 007 (2011).
[arXiv:1101.0557 [hep-th]].
}

\lref\ImamuraUW{
  Y.~Imamura,
 ``Relation between the 4d superconformal index and the $S^3$ partition function,''
JHEP {\bf 1109}, 133 (2011).
[arXiv:1104.4482 [hep-th]].
}

\lref\ImamuraWG{
  Y.~Imamura and D.~Yokoyama,
 ``N=2 supersymmetric theories on squashed three-sphere,''
Phys.\ Rev.\ D {\bf 85}, 025015 (2012).
[arXiv:1109.4734 [hep-th]].
}

\lref\ImamuraRQ{
  Y.~Imamura and D.~Yokoyama,
 ``$S^3/Z_n$ partition function and dualities,''
JHEP {\bf 1211}, 122 (2012).
[arXiv:1208.1404 [hep-th]].
}

\lref\InbasekarTSA{
  K.~Inbasekar, S.~Jain, S.~Mazumdar, S.~Minwalla, V.~Umesh and S.~Yokoyama,
  ``Unitarity, crossing symmetry and duality in the scattering of ${\cal N}{=}1 $ SUSY matter Chern-Simons theories,''
JHEP {\bf 1510}, 176 (2015).
[arXiv:1505.06571 [hep-th]].
}

\lref\IntriligatorID{
  K.~A.~Intriligator and N.~Seiberg,
  ``Duality, monopoles, dyons, confinement and oblique confinement in supersymmetric SO(N(c)) gauge theories,''
Nucl.\ Phys.\ B {\bf 444}, 125 (1995).
[hep-th/9503179].
}

\lref\IntriligatorNE{
  K.~A.~Intriligator and P.~Pouliot,
  ``Exact superpotentials, quantum vacua and duality in supersymmetric SP(N(c)) gauge theories,''
Phys.\ Lett.\ B {\bf 353}, 471 (1995).
[hep-th/9505006].
}

\lref\IntriligatorER{
  K.~A.~Intriligator and N.~Seiberg,
  ``Phases of N=1 supersymmetric gauge theories and electric - magnetic triality,''
In *Los Angeles 1995, Future perspectives in string theory* 270-282.
[hep-th/9506084].
}

\lref\IntriligatorAU{
  K.~A.~Intriligator and N.~Seiberg,
  ``Lectures on supersymmetric gauge theories and electric - magnetic duality,''
Nucl.\ Phys.\ Proc.\ Suppl.\  {\bf 45BC}, 1 (1996).
[hep-th/9509066].
}

\lref\IntriligatorEX{
  K.~A.~Intriligator and N.~Seiberg,
  ``Mirror symmetry in three-dimensional gauge theories,''
Phys.\ Lett.\ B {\bf 387}, 513 (1996).
[hep-th/9607207].
}

\lref\IntriligatorLCA{
  K.~Intriligator and N.~Seiberg,
  ``Aspects of 3d ${\cal N}{=}2$ Chern-Simons-Matter Theories,''
JHEP {\bf 1307}, 079 (2013).
[arXiv:1305.1633 [hep-th]].
}

\lref\IvanovFN{
   E.~A.~Ivanov,
   ``Chern-Simons matter systems with manifest N=2 supersymmetry,''
Phys.\ Lett.\ B {\bf 268}, 203 (1991).
}

\lref\JafferisUN{
  D.~L.~Jafferis,
  ``The Exact Superconformal R-Symmetry Extremizes Z,''
JHEP {\bf 1205}, 159 (2012).
[arXiv:1012.3210 [hep-th]].
}

\lref\JafferisZI{
  D.~L.~Jafferis, I.~R.~Klebanov, S.~S.~Pufu and B.~R.~Safdi,
  ``Towards the F-Theorem: N=2 Field Theories on the Three-Sphere,''
JHEP {\bf 1106}, 102 (2011).
[arXiv:1103.1181 [hep-th]].
}

\lref\JainTX{
  J.~K.~Jain,
  ``Composite fermion approach for the fractional quantum Hall effect,''
Phys.\ Rev.\ Lett.\  {\bf 63}, 199 (1989).
}

\lref\JainPY{
  S.~Jain, S.~Minwalla, T.~Sharma, T.~Takimi, S.~R.~Wadia and S.~Yokoyama,
  ``Phases of large $N$ vector Chern-Simons theories on $S^2 {\times} S^1$,''
JHEP {\bf 1309}, 009 (2013).
[arXiv:1301.6169 [hep-th]].
}

\lref\JainGZA{
  S.~Jain, S.~Minwalla and S.~Yokoyama,
  ``Chern Simons duality with a fundamental boson and fermion,''
JHEP {\bf 1311}, 037 (2013).
[arXiv:1305.7235 [hep-th]].
}

\lref\JainNZA{
  S.~Jain, M.~Mandlik, S.~Minwalla, T.~Takimi, S.~R.~Wadia and S.~Yokoyama,
  ``Unitarity, Crossing Symmetry and Duality of the S-matrix in large $N$ Chern-Simons theories with fundamental matter,''
JHEP {\bf 1504}, 129 (2015).
[arXiv:1404.6373 [hep-th]].
}

\lref\KachruRMA{
  S.~Kachru, M.~Mulligan, G.~Torroba and H.~Wang,
  ``Mirror symmetry and the half-filled Landau level,''
Phys.\ Rev.\ B {\bf 92}, 235105 (2015).
[arXiv:1506.01376 [cond-mat.str-el]].
}

\lref\KachruRUI{
  S.~Kachru, M.~Mulligan, G.~Torroba and H.~Wang,
  ``Bosonization and Mirror Symmetry,''
Phys.\ Rev.\ D {\bf 94}, no. 8, 085009 (2016).
[arXiv:1608.05077 [hep-th]].
}

\lref\KachruAON{
  S.~Kachru, M.~Mulligan, G.~Torroba and H.~Wang,
  ``Nonsupersymmetric dualities from mirror symmetry,''
Phys.\ Rev.\ Lett.\  {\bf 118}, 011602 (2017).
[arXiv:1609.02149 [hep-th]].
}

\lref\KajantieVY{
  K.~Kajantie, M.~Laine, T.~Neuhaus, A.~Rajantie and K.~Rummukainen,
  ``Duality and scaling in three-dimensional scalar electrodynamics,''
Nucl.\ Phys.\ B {\bf 699}, 632 (2004).
[hep-lat/0402021].
}

\lref\KapustinHA{
  A.~Kapustin and M.~J.~Strassler,
  ``On mirror symmetry in three-dimensional Abelian gauge theories,''
JHEP {\bf 9904}, 021 (1999).
[hep-th/9902033].
}

\lref\KapustinPY{
  A.~Kapustin,
  ``Wilson-'t Hooft operators in four-dimensional gauge theories and S-duality,''
Phys.\ Rev.\ D {\bf 74}, 025005 (2006).
[hep-th/0501015].
}

\lref\KapustinKZ{
  A.~Kapustin, B.~Willett and I.~Yaakov,
  ``Exact Results for Wilson Loops in Superconformal Chern-Simons Theories with Matter,''
JHEP {\bf 1003}, 089 (2010).
[arXiv:0909.4559 [hep-th]].
}

\lref\KapustinSim{
A.~Kapustin,  2010 Simons Workshop talk, a video of this talk can be found at
{\tt
http://media.scgp.stonybrook.edu/video/video.php?f=20110810\_1\_qtp.mp4}
}

\lref\KapustinXQ{
  A.~Kapustin, B.~Willett and I.~Yaakov,
  ``Nonperturbative Tests of Three-Dimensional Dualities,''
JHEP {\bf 1010}, 013 (2010).
[arXiv:1003.5694 [hep-th]].
}

\lref\KapustinGH{
  A.~Kapustin,
  ``Seiberg-like duality in three dimensions for orthogonal gauge groups,''
arXiv:1104.0466 [hep-th].
}

\lref\KapustinJM{
  A.~Kapustin and B.~Willett,
  ``Generalized Superconformal Index for Three Dimensional Field Theories,''
[arXiv:1106.2484 [hep-th]].
}

\lref\KapustinVZ{
  A.~Kapustin, H.~Kim and J.~Park,
  ``Dualities for 3d Theories with Tensor Matter,''
JHEP {\bf 1112}, 087 (2011).
[arXiv:1110.2547 [hep-th]].
}

\lref\KapustinGUA{
  A.~Kapustin and N.~Seiberg,
  ``Coupling a QFT to a TQFT and Duality,''
JHEP {\bf 1404}, 001 (2014).
[arXiv:1401.0740 [hep-th]].
}

\lref\KarchUX{
  A.~Karch,
  ``Seiberg duality in three-dimensions,''
Phys.\ Lett.\ B {\bf 405}, 79 (1997).
[hep-th/9703172].
}

\lref\KarchSXI{
  A.~Karch and D.~Tong,
  ``Particle-Vortex Duality from 3d Bosonization,''
Phys.\ Rev.\ X {\bf 6}, 031043 (2016). 
[arXiv:1606.01893 [hep-th]].
}

\lref\KarchAUX{
  A.~Karch, B.~Robinson and D.~Tong,
  ``More Abelian Dualities in 2+1 Dimensions,''
JHEP {\bf 1701}, 017 (2017).
[arXiv:1609.04012 [hep-th]].
}

\lref\KimWB{
  S.~Kim,
  ``The Complete superconformal index for N=6 Chern-Simons theory,''
Nucl.\ Phys.\ B {\bf 821}, 241 (2009), [Erratum-ibid.\ B {\bf 864}, 884 (2012)].
[arXiv:0903.4172 [hep-th]].
}

\lref\KimCMA{
  H.~Kim and J.~Park,
  ``Aharony Dualities for 3d Theories with Adjoint Matter,''
[arXiv:1302.3645 [hep-th]].
}

\lref\KinneyEJ{
  J.~Kinney, J.~M.~Maldacena, S.~Minwalla and S.~Raju,
  ``An Index for 4 dimensional super conformal theories,''
Commun.\ Math.\ Phys.\  {\bf 275}, 209 (2007).
[hep-th/0510251].
}

\lref\KlebanovJA{
  I.~R.~Klebanov and A.~M.~Polyakov,
  ``AdS dual of the critical $O(N)$ vector model,''
Phys.\ Lett.\ B {\bf 550}, 213 (2002).
[hep-th/0210114].
}

\lref\KrattenthalerDA{
  C.~Krattenthaler, V.~P.~Spiridonov, G.~S.~Vartanov,
  ``Superconformal indices of three-dimensional theories related by mirror symmetry,''
JHEP {\bf 1106}, 008 (2011).
[arXiv:1103.4075 [hep-th]].
}

\lref\McG{
  S. M. Kravec, J. McGreevy, and B. Swingle,
  ``All-Fermion Electrodynamics And Fermion Number Anomaly Inflow,''
arXiv:1409.8339.
}

\lref\KutasovVE{
  D.~Kutasov,
  ``A Comment on duality in N=1 supersymmetric nonAbelian gauge theories,''
Phys.\ Lett.\ B {\bf 351}, 230 (1995).
[hep-th/9503086].
}

\lref\KutasovNP{
  D.~Kutasov and A.~Schwimmer,
  ``On duality in supersymmetric Yang-Mills theory,''
Phys.\ Lett.\ B {\bf 354}, 315 (1995).
[hep-th/9505004].
}

\lref\KutasovSS{
  D.~Kutasov, A.~Schwimmer and N.~Seiberg,
  ``Chiral rings, singularity theory and electric - magnetic duality,''
Nucl.\ Phys.\ B {\bf 459}, 455 (1996).
[hep-th/9510222].
}

\lref\LeeVP{
  K.~-M.~Lee and P.~Yi,
  ``Monopoles and instantons on partially compactified D-branes,''
Phys.\ Rev.\ D {\bf 56}, 3711 (1997).
[hep-th/9702107].
}

\lref\LeeVU{
  K.~-M.~Lee,
  ``Instantons and magnetic monopoles on R**3 x S**1 with arbitrary simple gauge groups,''
Phys.\ Lett.\ B {\bf 426}, 323 (1998).
[hep-th/9802012].
}

\lref\MaldacenaSS{
  J.~M.~Maldacena, G.~W.~Moore and N.~Seiberg,
  ``D-brane charges in five-brane backgrounds,''
JHEP {\bf 0110}, 005 (2001).
[hep-th/0108152].
}

\lref\MaldacenaJN{
  J.~Maldacena and A.~Zhiboedov,
  ``Constraining Conformal Field Theories with A Higher Spin Symmetry,''
J.\ Phys.\ A {\bf 46}, 214011 (2013).
[arXiv:1112.1016 [hep-th]].
}

\lref\MetlitskiEKA{
  M.~A.~Metlitski and A.~Vishwanath,
  ``Particle-vortex duality of 2d Dirac fermion from electric-magnetic duality of 3d topological insulators,''
[arXiv:1505.05142 [cond-mat.str-el]].
}

\lref\MetlitskiBPA{
  M.~A.~Metlitski, C.~L.~Kane and M.~P.~A.~Fisher,
  ``Symmetry-respecting topologically ordered surface phase of three-dimensional electron topological insulators,''
Phys.\ Rev.\ B {\bf 92}, no. 12, 125111 (2015).
}

\lref\MetlitskiYQA{
  M.~A.~Metlitski,
  ``$S$-duality of $u(1)$ gauge theory with $\theta =\pi$ on non-orientable manifolds: Applications to topological insulators and superconductors,''
[arXiv:1510.05663 [hep-th]].
}

\lref\MetlitskiDHT{
  M.~A.~Metlitski, A.~Vishwanath and C.~Xu,
  ``Duality and bosonization of (2+1)d Majorana fermions,''
  arXiv:1611.05049 [cond-mat.str-el].
}

\lref\MinwallaSCA{
  S.~Minwalla and S.~Yokoyama,
  ``Chern Simons Bosonization along RG Flows,''
JHEP {\bf 1602}, 103 (2016).
[arXiv:1507.04546 [hep-th]].
}

\lref\MlawerUV{
  E.~J.~Mlawer, S.~G.~Naculich, H.~A.~Riggs and H.~J.~Schnitzer,
  ``Group level duality of WZW fusion coefficients and Chern-Simons link observables,''
Nucl.\ Phys.\ B {\bf 352}, 863 (1991).
}

\lref\MSN{
G.~W.~Moore and N.~Seiberg,
  ``Naturality in Conformal Field Theory,''
  Nucl.\ Phys.\ B {\bf 313}, 16 (1989).}

\lref\MooreYH{
  G.~W.~Moore and N.~Seiberg,
  ``Taming the Conformal Zoo,''
Phys.\ Lett.\ B {\bf 220}, 422 (1989).
}

\lref\MoritaCS{
  T.~Morita and V.~Niarchos,
  ``F-theorem, duality and SUSY breaking in one-adjoint Chern-Simons-Matter theories,''
Nucl.\ Phys.\ B {\bf 858}, 84 (2012).
[arXiv:1108.4963 [hep-th]].
}

\lref\MrossIDY{
  D.~F.~Mross, J.~Alicea and O.~I.~Motrunich,
  ``Explicit derivation of duality between a free Dirac cone and quantum electrodynamics in (2+1) dimensions,''
[arXiv:1510.08455 [cond-mat.str-el]].
}

\lref\MulliganGLM{
  M.~Mulligan, S.~Raghu and M.~P.~A.~Fisher,
  ``Emergent particle-hole symmetry in the half-filled Landau level,''
[arXiv:1603.05656 [cond-mat.str-el]].
}

\lref\MuruganZAL{
  J.~Murugan and H.~Nastase,
  ``Particle-vortex duality in topological insulators and superconductors,''
[arXiv:1606.01912 [hep-th]].
}

\lref\NaculichPA{
  S.~G.~Naculich, H.~A.~Riggs and H.~J.~Schnitzer,
  ``Group Level Duality in {WZW} Models and {Chern-Simons} Theory,''
Phys.\ Lett.\ B {\bf 246}, 417 (1990).
}

\lref\NaculichNC{
  S.~G.~Naculich and H.~J.~Schnitzer,
  ``Level-rank duality of the U(N) WZW model, Chern-Simons theory, and 2-D qYM theory,''
JHEP {\bf 0706}, 023 (2007).
[hep-th/0703089 [HEP-TH]].
}

\lref\NakaharaNW{
  M.~Nakahara,
  ``Geometry, topology and physics,''
Boca Raton, USA: Taylor and Francis (2003) 573 p.
}

\lref\NakanishiHJ{
  T.~Nakanishi and A.~Tsuchiya,
  ``Level rank duality of WZW models in conformal field theory,''
Commun.\ Math.\ Phys.\  {\bf 144}, 351 (1992).
}

\lref\NguyenZN{
  A.~K.~Nguyen and A.~Sudbo,
  ``Topological phase fluctuations, amplitude fluctuations, and criticality in extreme type II superconductors,''
Phys.\ Rev.\ B {\bf 60}, 15307 (1999).
[cond-mat/9907385].
}

\lref\NiarchosJB{
  V.~Niarchos,
  ``Seiberg Duality in Chern-Simons Theories with Fundamental and Adjoint Matter,''
JHEP {\bf 0811}, 001 (2008).
[arXiv:0808.2771 [hep-th]].
}

\lref\NiarchosAA{
  V.~Niarchos,
  ``R-charges, Chiral Rings and RG Flows in Supersymmetric Chern-Simons-Matter Theories,''
JHEP {\bf 0905}, 054 (2009).=
[arXiv:0903.0435 [hep-th]].
}

\lref\NiarchosAH{
  V.~Niarchos,
  ``Seiberg dualities and the 3d/4d connection,''
JHEP {\bf 1207}, 075 (2012).
[arXiv:1205.2086 [hep-th]].
}

\lref\NiemiRQ{
  A.~J.~Niemi and G.~W.~Semenoff,
  ``Axial Anomaly Induced Fermion Fractionization and Effective Gauge Theory Actions in Odd Dimensional Space-Times,''
Phys.\ Rev.\ Lett.\  {\bf 51}, 2077 (1983).
}

\lref\VOstrik{
  V.~Ostrik and M.~Sun,
  ``Level-Rank Duality Via Tensor Categories,''
Comm. Math. Phys. 326 (2014) 49-61.
[arXiv:1208.5131 [math-ph]].
}

\lref\ParkWTA{
  J.~Park and K.~J.~Park,
  ``Seiberg-like Dualities for 3d ${\cal N}{=}2$ Theories with $SU(N)$ gauge group,''
JHEP {\bf 1310}, 198 (2013).
[arXiv:1305.6280 [hep-th]].
}

\lref\PaulyAMA{
  C.~Pauly,
  ``Strange duality revisited,''
Math.\ Res.\ Lett.\  {\bf 21}, 1353 (2014).
}

\lref\PeskinKP{
  M.~E.~Peskin,
  ``Mandelstam 't Hooft Duality in Abelian Lattice Models,''
Annals Phys.\  {\bf 113}, 122 (1978).
}

\lref\PolyakovFU{
  A.~M.~Polyakov,
  ``Quark Confinement and Topology of Gauge Groups,''
Nucl.\ Phys.\ B {\bf 120}, 429 (1977).
}

\lref\PolyakovMD{
  A.~M.~Polyakov,
  ``Fermi-Bose Transmutations Induced by Gauge Fields,''
Mod.\ Phys.\ Lett.\ A {\bf 3}, 325 (1988).
}

\lref\PotterCDN{
  A.~C.~Potter, M.~Serbyn and A.~Vishwanath,
  ``Thermoelectric transport signatures of Dirac composite fermions in the half-filled Landau level,''
Phys.\ Rev.\ X {\bf 6}, 031026 (2016).
[arXiv:1512.06852 [cond-mat.str-el]].
}

\lref\QiEW{
  X.~L.~Qi, T.~Hughes and S.~C.~Zhang,
  ``Topological Field Theory of Time-Reversal Invariant Insulators,''
Phys.\ Rev.\ B {\bf 78}, 195424 (2008).
[arXiv:0802.3537 [cond-mat.mes-hall]].
}

\lref\CordovaVAB{
  C.~Cordova, P.~S.~Hsin and N.~Seiberg,
  ``Global Symmetries, Counterterms, and Duality in Chern-Simons Matter Theories with Orthogonal Gauge Groups,''
[arXiv:1711.10008 [hep-th]].
}

\lref\RabinoviciMJ{
  E.~Rabinovici, A.~Schwimmer and S.~Yankielowicz,
  ``Quantization in the Presence of {Wess-Zumino} Terms,''
Nucl.\ Phys.\ B {\bf 248}, 523 (1984).
}

\lref\RadicevicYLA{
  D.~Radicevic,
  ``Disorder Operators in Chern-Simons-Fermion Theories,''
JHEP {\bf 1603}, 131 (2016).
[arXiv:1511.01902 [hep-th]].
}

\lref\RadicevicWQN{
  D.~Radicevic, D.~Tong and C.~Turner,
  ``Non-Abelian 3d Bosonization and Quantum Hall States,''
JHEP {\bf 1612}, 067 (2016).
[arXiv:1608.04732 [hep-th]].
}

\lref\RazamatUV{
  S.~S.~Razamat,
  ``On a modular property of N=2 superconformal theories in four dimensions,''
JHEP {\bf 1210}, 191 (2012).
[arXiv:1208.5056 [hep-th]].
}

\lref\RedlichDV{
  A.~N.~Redlich,
  ``Parity Violation and Gauge Noninvariance of the Effective Gauge Field Action in Three-Dimensions,''
Phys.\ Rev.\ D {\bf 29}, 2366 (1984).
}

\lref\Rehren{
  K.-H.~Rehren,
  ``Algebraic Conformal QFT'',
3rd Meeting of the French-Italian Research Team on Noncommutative Geometry and Quantum Physics Vietri sul Mare, 2009.
}

\lref\RomelsbergerEG{
  C.~Romelsberger,
  ``Counting chiral primaries in N = 1, d=4 superconformal field theories,''
Nucl.\ Phys.\ B {\bf 747}, 329 (2006).
[hep-th/0510060].
}

\lref\RoscherWOX{
  D.~Roscher, E.~Torres and P.~Strack,
  ``Dual QED$_3$ at "$N_F = 1/2$" is an interacting CFT in the infrared,''
[arXiv:1605.05347 [cond-mat.str-el]].
}

\lref\SafdiRE{
  B.~R.~Safdi, I.~R.~Klebanov and J.~Lee,
  ``A Crack in the Conformal Window,''
[arXiv:1212.4502 [hep-th]].
}

\lref\SeibergBZ{
  N.~Seiberg,
  ``Exact results on the space of vacua of four-dimensional SUSY gauge theories,''
Phys.\ Rev.\ D {\bf 49}, 6857 (1994).
[hep-th/9402044].
}

\lref\SeibergPQ{
  N.~Seiberg,
  ``Electric - magnetic duality in supersymmetric nonAbelian gauge theories,''
Nucl.\ Phys.\ B {\bf 435}, 129 (1995).
[hep-th/9411149].
}

\lref\SeibergNZ{
  N.~Seiberg and E.~Witten,
  ``Gauge dynamics and compactification to three-dimensions,''
In *Saclay 1996, The mathematical beauty of physics* 333-366.
[hep-th/9607163].
}

\lref\SeibergQD{
  N.~Seiberg,
  ``Modifying the Sum Over Topological Sectors and Constraints on Supergravity,''
JHEP {\bf 1007}, 070 (2010).
[arXiv:1005.0002 [hep-th]].
}

\lref\SeibergRSG{
  N.~Seiberg and E.~Witten,
  ``Gapped Boundary Phases of Topological Insulators via Weak Coupling,''
PTEP {\bf 2016}, 12C101 (2016). 
[arXiv:1602.04251 [cond-mat.str-el]].
}

\lref\SeibergGMD{
  N.~Seiberg, T.~Senthil, C.~Wang and E.~Witten,
  ``A Duality Web in 2+1 Dimensions and Condensed Matter Physics,''
Annals Phys.\  {\bf 374}, 395 (2016).
[arXiv:1606.01989 [hep-th]].
}

\lref\SenthilJK{
  T.~Senthil and M.~P.~A.~Fisher,
  ``Competing orders, non-linear sigma models, and topological terms in quantum magnets,''
Phys.\ Rev.\ B {\bf 74}, 064405 (2006).
[cond-mat/0510459].
}

\lref\SezginRT{
  E.~Sezgin and P.~Sundell,
  ``Massless higher spins and holography,''
Nucl.\ Phys.\ B {\bf 644}, 303 (2002), Erratum: [Nucl.\ Phys.\ B {\bf 660}, 403 (2003)].
[hep-th/0205131].
}

\lref\SezginPT{
  E.~Sezgin and P.~Sundell,
  ``Holography in 4D (super) higher spin theories and a test via cubic scalar couplings,''
JHEP {\bf 0507}, 044 (2005).
[hep-th/0305040].
}

\lref\ShajiIS{
  N.~Shaji, R.~Shankar and M.~Sivakumar,
  ``On Bose-fermi Equivalence in a U(1) Gauge Theory With {Chern-Simons} Action,''
Mod.\ Phys.\ Lett.\ A {\bf 5}, 593 (1990).
}

\lref\Shamirthesis{
  I.~Shamir,
  ``Aspects of three dimensional Seiberg duality,''
  M.~Sc. thesis submitted to the Weizmann Institute of Science, April 2010.
}

\lref\ShenkerZF{
  S.~H.~Shenker and X.~Yin,
  ``Vector Models in the Singlet Sector at Finite Temperature,''
[arXiv:1109.3519 [hep-th]].
}

\lref\SonXQA{
  D.~T.~Son,
  ``Is the Composite Fermion a Dirac Particle?,''
Phys.\ Rev.\ X {\bf 5}, 031027 (2015).
[arXiv:1502.03446 [cond-mat.mes-hall]].
}

\lref\SpiridonovZR{
  V.~P.~Spiridonov and G.~S.~Vartanov,
  ``Superconformal indices for N = 1 theories with multiple duals,''
Nucl.\ Phys.\ B {\bf 824}, 192 (2010).
[arXiv:0811.1909 [hep-th]].
}

\lref\SpiridonovZA{
  V.~P.~Spiridonov and G.~S.~Vartanov,
  ``Elliptic Hypergeometry of Supersymmetric Dualities,''
Commun.\ Math.\ Phys.\  {\bf 304}, 797 (2011).
[arXiv:0910.5944 [hep-th]].
}

\lref\SpiridonovHF{
  V.~P.~Spiridonov and G.~S.~Vartanov,
  ``Elliptic hypergeometry of supersymmetric dualities II. Orthogonal groups, knots, and vortices,''
[arXiv:1107.5788 [hep-th]].
}

\lref\SpiridonovWW{
  V.~P.~Spiridonov and G.~S.~Vartanov,
  ``Elliptic hypergeometric integrals and 't Hooft anomaly matching conditions,''
JHEP {\bf 1206}, 016 (2012).
[arXiv:1203.5677 [hep-th]].
}

\lref\StrasslerFE{
  M.~J.~Strassler,
  ``Duality, phases, spinors and monopoles in $SO(N)$ and $spin(N)$ gauge theories,''
JHEP {\bf 9809}, 017 (1998).
[hep-th/9709081].
}

\lref\VasilievVF{
  M.~A.~Vasiliev,
  ``Holography, Unfolding and Higher-Spin Theory,''
J.\ Phys.\ A {\bf 46}, 214013 (2013).
[arXiv:1203.5554 [hep-th]].
}

\lref\VerstegenAT{
  D.~Verstegen,
  ``Conformal embeddings, rank-level duality and exceptional modular invariants,''
Commun.\ Math.\ Phys.\  {\bf 137}, 567 (1991).
}

\lref\WangUKY{
  C.~Wang, A.~C.~Potter and T.~Senthil,
  ``Gapped symmetry preserving surface state for the electron topological insulator,''
Phys.\ Rev.\ B {\bf 88}, no. 11, 115137 (2013).
[arXiv:1306.3223 [cond-mat.str-el]].
}

\lref\MPS{
  C. Wang, A. C. Potter, and T. Senthil,
  ``Classification Of Interacting Electronic Topological Insulators In Three Dimensions,''
Science {\bf 343} (2014) 629,
[arXiv:1306.3238].
}

\lref\WangLCA{
  C.~Wang and T.~Senthil,
  ``Interacting fermionic topological insulators/superconductors in three dimensions,''
Phys.\ Rev.\ B {\bf 89}, no. 19, 195124 (2014), Erratum: [Phys.\ Rev.\ B {\bf 91}, no. 23, 239902 (2015)].
[arXiv:1401.1142 [cond-mat.str-el]].
}

\lref\WangQMT{
  C.~Wang and T.~Senthil,
  ``Dual Dirac Liquid on the Surface of the Electron Topological Insulator,''
Phys.\ Rev.\ X {\bf 5}, no. 4, 041031 (2015). [arXiv:1505.05141 [cond-mat.str-el]].
}

\lref\WangFQL{
  C.~Wang and T.~Senthil,
  ``Half-filled Landau level, topological insulator surfaces, and three-dimensional quantum spin liquids,''
Phys.\ Rev.\ B {\bf 93}, no. 8, 085110 (2016). [arXiv:1507.08290 [cond-mat.st-el]].
}

\lref\WangGQJ{
  C.~Wang and T.~Senthil,
  ``Composite fermi liquids in the lowest Landau level,''
Phys.\ Rev.\ B {\bf 94}, 245107 (2016). 
[arXiv:1604.06807 [cond-mat.str-el]].
}

\lref\WangCTO{
  C.~Wang and T.~Senthil,
  ``Time-Reversal Symmetric $U(1)$ Quantum Spin Liquids,''
Phys.\ Rev.\ X {\bf 6}, no. 1, 011034 (2016).
}

\lref\Whitehead{
  J.~H.~C.~Whitehead,
  ``On simply connected, 4-dimensional polyhedra,''
Comm.\ Math.\ Helv.\ {\bf 22} (1949) 48.
}

\lref\WilczekDU{
  F.~Wilczek,
  ``Magnetic Flux, Angular Momentum, and Statistics,''
Phys.\ Rev.\ Lett.\  {\bf 48}, 1144 (1982).
}

\lref\WilczekCY{
  F.~Wilczek and A.~Zee,
  ``Linking Numbers, Spin, and Statistics of Solitons,''
Phys.\ Rev.\ Lett.\  {\bf 51}, 2250 (1983).
}

\lref\GasserYG{
  J.~Gasser and H.~Leutwyler,
  ``Chiral Perturbation Theory to One Loop,''
Annals Phys.\  {\bf 158}, 142 (1984).
}

\lref\WillettGP{
  B.~Willett and I.~Yaakov,
  ``${\cal N}{=}2$ Dualities and $Z$-extremization in Three Dimensions,''
arXiv:1104.0487 [hep-th].
}

\lref\WittenHF{
  E.~Witten,
  ``Quantum Field Theory and the Jones Polynomial,''
Commun.\ Math.\ Phys.\  {\bf 121}, 351 (1989).
}

\lref\ClossetVP{
  C.~Closset, T.~T.~Dumitrescu, G.~Festuccia, Z.~Komargodski and N.~Seiberg,
  ``Comments on Chern-Simons Contact Terms in Three Dimensions,''
JHEP {\bf 1209}, 091 (2012).
[arXiv:1206.5218 [hep-th]].
}

\lref\WittenXI{
  E.~Witten,
  ``The Verlinde algebra and the cohomology of the Grassmannian,''
In *Cambridge 1993, Geometry, topology, and physics* 357-422.
[hep-th/9312104].
}

\lref\WittenGF{
  E.~Witten,
  ``On S duality in Abelian gauge theory,''
Selecta Math.\  {\bf 1}, 383 (1995).
[hep-th/9505186].
}

\lref\WittenDS{
  E.~Witten,
  ``Supersymmetric index of three-dimensional gauge theory,''
In *Shifman, M.A. (ed.): The many faces of the superworld* 156-184.
[hep-th/9903005].
}

\lref\DiPietroKCD{
  L.~Di Pietro and E.~Stamou,
  ``Scaling dimensions in QED$_3$ from the $\epsilon$-expansion,''
[arXiv:1708.03740 [hep-th]].
}

\lref\DiPietroVSP{
  L.~Di Pietro and E.~Stamou,
  ``Operator mixing in $\epsilon$-expansion: scheme and evanescent (in)dependence,''
[arXiv:1708.03739 [hep-th]].
}

\lref\DiPietroYKH{
  L.~Di Pietro, Z.~Komargodski, I.~Shamir and E.~Stamou,
  ``The $\epsilon$ -expansion for QED,''
Int.\ J.\ Mod.\ Phys.\ A {\bf 31}, no. 28\&29, 1645039 (2016)..
}

\lref\WittenYA{
  E.~Witten,
  ``SL(2,Z) action on three-dimensional conformal field theories with Abelian symmetry,''
In *Shifman, M. (ed.) et al.: From fields to strings, vol. 2* 1173-1200.
[hep-th/0307041].
}

\lref\WittenABA{
  E.~Witten,
  ``Fermion Path Integrals And Topological Phases,''
[arXiv:1508.04715 [cond-mat.mes-hall]].
}

\lref\WuGE{
  T.~T.~Wu and C.~N.~Yang,
  ``Dirac Monopole Without Strings: Monopole Harmonics,''
Nucl.\ Phys.\ B {\bf 107}, 365 (1976).
}

\lref\XuNXA{
  F.~Xu,
  ``Algebraic coset conformal field theories,''
Commun.\ Math.\ Phys.\  {\bf 211}, 1 (2000).
[math/9810035].
}

\lref\VafaTF{
  C.~Vafa and E.~Witten,
  ``Restrictions on Symmetry Breaking in Vector-Like Gauge Theories,''
Nucl.\ Phys.\ B {\bf 234}, 173 (1984).
}

\lref\BeniniDUS{
  F.~Benini, P.~S.~Hsin and N.~Seiberg,
  ``Comments on global symmetries, anomalies, and duality in (2 + 1)d,''
JHEP {\bf 1704}, 135 (2017).
[arXiv:1702.07035 [cond-mat.str-el]].
}

\lref\RedlichDV{
  A.~N.~Redlich,
  ``Parity Violation and Gauge Noninvariance of the Effective Gauge Field Action in Three-Dimensions,''
Phys.\ Rev.\ D {\bf 29}, 2366 (1984).
}

\lref\XuLXA{
  C.~Xu and Y.~Z.~You,
  ``Self-dual Quantum Electrodynamics as Boundary State of the three dimensional Bosonic Topological Insulator,''
Phys.\ Rev.\ B {\bf 92}, 220416 (2015). 
[arXiv:1510.06032 [cond-mat.str-el]].
}

\lref\RadicevicYLA{
  D.~Radicevic,
  ``Disorder Operators in Chern-Simons-Fermion Theories,''
JHEP {\bf 1603}, 131 (2016).
[arXiv:1511.01902 [hep-th]].
}

\lref\AharonyMJS{
  O.~Aharony,
  ``Baryons, monopoles and dualities in Chern-Simons-matter theories,''
JHEP {\bf 1602}, 093 (2016).
[arXiv:1512.00161 [hep-th]].
}

\lref\ZupnikRY{
   B.~M.~Zupnik and D.~G.~Pak,
   ``Topologically Massive Gauge Theories In Superspace,''
Sov.\ Phys.\ J.\  {\bf 31}, 962 (1988).
}

\lref\AppelquistTC{
  T.~Appelquist and D.~Nash,
  ``Critical Behavior in (2+1)-dimensional {QCD},''
Phys.\ Rev.\ Lett.\  {\bf 64}, 721 (1990).
}

\lref\FreedMX{
  D.~S.~Freed,
  ``Pions and Generalized Cohomology,''
J.\ Diff.\ Geom.\  {\bf 80}, no. 1, 45 (2008).
[hep-th/0607134].
}

\lref\ZwiebelWA{
  B.~I.~Zwiebel,
  ``Charging the Superconformal Index,''
JHEP {\bf 1201}, 116 (2012).
[arXiv:1111.1773 [hep-th]].
}

\lref\KomargodskiDMC{
  Z.~Komargodski, A.~Sharon, R.~Thorngren and X.~Zhou,
  ``Comments on Abelian Higgs Models and Persistent Order,''
[arXiv:1705.04786 [hep-th]].
}

\lref\PaperTwo{
  D.~Gaiotto, Z.~Komargodski and N.~Seiberg,
  ``Time-Reversal Breaking in QCD$_4$, Walls, and Dualities in 2+1 Dimensions,''
[arXiv:1708.06806 [hep-th]].
}

\lref\PaperThree{
  D.~S.~Freed, Z.~Komargodski and N.~Seiberg,
  ``The Sum Over Topological Sectors and $\theta$ in the 2+1-Dimensional $\C\P^1$ $\sigma$-Model,''
[arXiv:1707.05448 [cond-mat.str-el]].
}


%
%

\vskip-60pt
\Title{} {\vbox{\centerline{A Symmetry Breaking Scenario for QCD$_3$
}}  }

\vskip-15pt

\centerline{Zohar Komargodski${}^{1,2}$ and Nathan Seiberg${}^3$}
\vskip15pt
\centerline{\it ${}^1$ Department of Particle Physics and Astrophysics, Weizmann Institute of Science, Israel }
\centerline{\it ${}^2$   Simons Center for Geometry and Physics, Stony Brook University, Stony Brook, NY}
\centerline{\it ${}^3$ School of Natural Sciences, Institute for Advanced Study, Princeton, NJ 08540, USA}

\vskip25pt

\noindent
We consider the dynamics of 2+1 dimensional $SU(N)$ gauge theory with Chern-Simons level $k$ and $N_f$ fundamental fermions. By requiring consistency with previously suggested dualities for $N_f\leq 2k$ as well as the dynamics at $k=0$ we propose that the theory with $N_f> 2k$ breaks the $U(N_f)$ global symmetry spontaneously to  $U(N_f/2+k)\times U(N_f/2-k)$. In contrast to the 3+1 dimensional case, the symmetry breaking takes place in a range of quark masses and not just at one point. The target space never becomes parametrically large and the Nambu-Goldstone bosons are therefore not visible semi-classically. Such symmetry breaking is argued to take place in some intermediate range of the number of flavors, $2k< N_f< N_*(N,k)$, with the upper limit $N_*$ obeying various constraints. The Lagrangian for the Nambu-Goldstone bosons has to be supplemented by nontrivial Wess-Zumino terms that are necessary for the consistency of the picture, even at $k=0$. Furthermore, we suggest two scalar dual theories in this range of $N_f$.  A similar picture is developed for $SO(N)$ and $Sp(N)$ gauge theories.  It sheds new light on monopole condensation and confinement in the $SO(N)\ \&\  Spin(N)$ theories.

\bigskip
\Date{June 2017}


\newif\ifbf\bffalse
\let\BF=\bf
\def\bf{\bftrue\BF}


%
%

\newsec{Introduction}

In this paper we study QCD in 2+1 dimensions, namely non-Abelian $2+1$ dimensional gauge theories coupled to fermionic matter fields.  We will focus on $SU(N)$, $SO(N)$, and $Sp(N)$ gauge theories coupled to $N_f$ fermion flavors in the fundamental representation.  In $SU(N)$ each fermion flavor is in an $N$ dimensional complex representation, in $SO(N)$ each flavor is in an $N$ dimensional real representation, and in $Sp(N)$ each flavor is in a $2N$ dimensional pseudo-real representation. These are the 2+1 dimensional analogs of Quantum Chromodynamics. While there is a rather concrete picture for the dynamics of 3+1 dimensional QCD, the 2+1 dimensional counterpart is still poorly understood in part because these theories depend not just on the number of flavors and the gauge group but also on the Chern-Simons level. Our main goal is to propose a scenario for the phases of these theories as a function of the gauge group, Chern-Simons level $k$, and the number of flavors $N_f$.

We will follow the notation and conventions of \refs{\SeibergRSG\SeibergGMD\HsinBLU-\AharonyJVV}, which we summarize now. Let us start from a single Dirac fermion of charge 1 coupled to a gauge field $A$,
\eqn\calLd{{\cal L} = i \psi^\dagger \slash{D}_A \psi~.}
The effective action for the background field $A$ can be computed by performing the path integral over the fermion. The effective action for $A$ is however ambiguous, since we could add various counter-terms.  More precisely, we could add to the action
\eqn\kbaret{{k_{bare}\over 4\pi}\int A\wedge dA~,}
with $k_{bare}\in \Z$. We will fix this ambiguity by choosing the following convention for the effective action of the free fermion. If we add to the Lagrangian a positive mass term for the fermion, then our IR effective action has no Chern-Simons term for the background field. If, on the other hand, we add a negative fermion mass, then the IR effective action is $-{1\over 4\pi}\int A\wedge dA$, i.e.\ the Chern-Simons level in the infrared is $-1$. This convention for the effective action breaks time reversal symmetry, but this is unavoidable in 2+1 dimensions~\RedlichDV. Our specific convention amounts (up to a sign) to identifying the phase of the fermion path integral with the eta invariant (see~\WittenABA\ for a recent detailed discussion).
Our convention for fermions coupling to non-Abelian gauge fields is analogous.

With this convention, let us suppose that we start from $N_f$ fermions coupled to a $U(1)$ gauge field with Chern-Simons level $k_{bare}\in \Z$. Then, for positive mass the low energy effective theory contains a Chern-Simons term of level $k_{bare}$ and with negative mass the infrared Chern-Simons level is $k_{bare}-N_f$. We could label this theory by its bare Chern-Simons term and the number of matter particles. In order to conform with the common notation used in the literature we  define
$k=k_{bare}-N_f/2$, in terms of which the infrared levels are $k\pm N_f/2$ for positive and negative mass, respectively. This choice is more symmetric. Note that while $k_{bare}$ was an integer, $k$ may be a half integer, and it is a half integer if and only if the number of fermions is odd. Henceforth we label this theory by $U(1)_k+N_f$ fermions. Our labelling of theories with non-Abelian gauge groups is analogous.

In all the theories that we will discuss here, the bare Chern-Simons term must always be an integer.  Above, to define $k$, we shifted  the bare Chern-Simons term by $-N_f/2$. We can think\foot{In the general case of an interacting conformal field theory, the ``quantum'' Chern-Simons term was defined in~\refs{\ClossetVG,\ClossetVP}.  It can be an arbitrary real number.} about this shift as $k_{quantum}$ and then
\eqn\lablethe{k=k_{bare} +k_{quantum}~.}
In our conventions, $k_{quantum}=-N_f/2$ in the $SU(N)$, $SO(N)$, and $Sp(N)$ theories, and, for example, $k_{quantum}=-h/2$ in the $\CN=1$ supersymmetric theories with $h$ the dual Coxeter number ($N$ in $SU(N)$).
 When the fermions have masses and they are integrated out the Chern-Simons level in the low energy theory is
\eqn\klow{k_{IR}=k -\sign(m) k_{quantum} ~,}
which is a properly normalized Chern-Simons level because $k$ is a half integer if and only if $k_{quantum}$ is a half integer. Note that the theories with $k=0$ are distinguished, since they have manifest time reversal invariance at $m=0$.  Note also that weakly coupled bosons have $k_{quantum}=0$ in all cases.

Before we proceed, we would like to make a general comment about the effect of the global aspects of the gauge group on the dynamics of 2+1 dimensional systems.  Often our gauge symmetry $ G$ is connected but not simply connected and then the system has a magnetic global symmetry $H$.  The charged objects under that symmetry are local monopole operators.  One typical example is $ G=U(1)$, where the magnetic symmetry is $H=U(1)$, and another example, which we will discuss in section 3, is $G=SO(N)$, whose magnetic symmetry is $ H=\Z_2$.  Suppose we take a subgroup $\tilde H \subset  H$ where $\tilde H$ is isomorphic to some $\Z_n$. We couple $\tilde H$ to a dynamical $\Z_n$ gauge field. This has the effect of changing the gauge group $G$ to a multiple cover of it $\tilde G$ such that $\tilde G/ \tilde H = G$.\foot{Note that the magnetic global symmetry group $ H$ is always Abelian and here we gauge only a discrete subgroup of it, $\tilde H$.  If we gauge a continuous subgroup of $H$, then there are new local operators, monopole operators, and the discussion is more complicated.} For example, starting with $G=U(1)$ and $\tilde H=\Z_n$ the new gauge group is $\tilde G=U(1)$, but all the particle charges are multiplied by $n$.  Similarly, for $ G=SO(N)$  with $\tilde H = \Z_2$ the new gauge group is $\tilde G= Spin(N)$.

How does the dynamics of the $\tilde G$ theory differ from that of the original $ G $ gauge theory?  Since all we did to the $G $ gauge theory is to gauge a discrete cyclic global symmetry, the resulting $\tilde G$ gauge theory is an $\tilde H$ orbifold of the original theory.  It is important that in $2+1d$ such orbifolding does not change the phase diagram.  The phase transitions in the $G$ gauge theory are the same as in the $ \tilde G$ theory.  The only difference between the two theories is that the $\tilde G$ theory has fewer local monopole operators and it has more line operators from the twisted sector.   If the $G$ theory has a phase with some nonlinear sigma model (i.e.\ with a moduli space of vacua) on which $\tilde H$ acts, then the $\tilde G$ theory has the same phase and the target space of the nonlinear sigma model is the quotient of the original target space by $\tilde H$. We therefore see that the space of vacua could depend on global aspects of the gauge group. (This phenomenon does not occur in four dimensions.) This will be important in section 3.

As a simple example, consider the $U(1)_k$ gauge theory with $N_f$ fermions all with charge one.  This theory has a magnetic $H=U(1)$ global symmetry.  Let us quotient the theory by $\tilde H=\Z_n\subset U(1)$.  This is easily done in the Lagrangian by replacing the gauge field $a$ by $na$; i.e.\ all the fermions now have charge $n$ and the Chern-Simons level of the new theory is $kn^2$.  The reasoning above shows that the phase diagram and the phase transitions between them are independent of $n$.  In particular, the phase diagram of the $U(1)_0$ theory is the same as that of the $\R$ gauge theory without a Chern-Simons term~\IntriligatorLCA.
In section 3 we will consider a similar example with $SO(N)$ and $Spin(N)$ gauge theories.

The dynamics of 2+1 dimensional non-Abelian gauge theories has been recently revitalised by the proposal of boson/fermion dualities \refs{\AharonyMJS\KarchSXI-\MuruganZAL,\SeibergGMD,\HsinBLU,\RadicevicWQN\KachruRUI
\KachruAON\KarchAUX\MetlitskiDHT\AharonyJVV-\BeniniDUS}.  They were motivated by ideas in $2+1d$ field theory \refs{\PeskinKP\DasguptaZZ\BarkeshliIDA\SonXQA\PotterCDN-\WangGQJ},
supersymmetric quantum field theory \refs{\IntriligatorEX\deBoerMP\AharonyBX\AharonyGP\GiveonZN\KapustinGH
\WillettGP\BeniniMF\AharonyCI\IntriligatorLCA\AharonyDHA\ParkWTA-\AharonyKMA},
and string theory \refs{\SezginRT\KlebanovJA\GiombiYA\AharonyJZ\GiombiKC\MaldacenaJN\AharonyNH
\GiombiMS\AharonyNS\JainPY\JainGZA
\JainNZA\InbasekarTSA\MinwallaSCA-\GurAriXFF}.  Roughly speaking, these dualities shed light on theories with small $N_f$ not larger than the Chern-Simons level in the fermionic theory.  As stated, these dualities do not apply in the time-reversal invariant case $k=0$.  Below we will clarify the role of duality in this case.

The dualities can be summarized as
\eqn\dualities{\eqalign{
U(k+N_f/2)_N \ {\rm with}\ N_f\ {\rm scalars} \quad &\longleftrightarrow \quad SU(N)_{-k} \ {\rm with}\ N_f\ {\rm fermions} \cr
SO(k+N_f/2)_N \ {\rm with}\ N_f\ {\rm real\ scalars} \quad &\longleftrightarrow \quad SO(N)_{-k} \ {\rm with}\ N_f\ {\rm real\ fermions} \cr
Sp(k+N_f/2)_N \ {\rm with}\ N_f\ {\rm scalars} \quad &\longleftrightarrow \quad Sp(N)_{-k} \ {\rm with}\ N_f\ {\rm fermions}}}
and other dualities that follow from them.  (Our notation is $U(k)_N \equiv U(k)_{N,N}$.)  The unitary dualities were conjectured to hold for $\half N_f \leq k$, the symplectic dualities
for $\half N_f \leq k$, and the orthogonal dualities for $\half N_f \leq k-2$ if $N=1$, $\half N_f \leq k -1$ if $N=2$, and $\half N_f \leq k$ if $N>2$.
We will also need their time-reversed versions
\eqn\dualitiesi{\eqalign{
U(k+N_f/2)_{-N} \ {\rm with}\ N_f\ {\rm scalars} \quad &\longleftrightarrow \quad SU(N)_{k} \ {\rm with}\ N_f\ {\rm fermions} \cr
SO(k+N_f/ 2)_{-N} \ {\rm with}\ N_f\ {\rm real\ scalars} \quad &\longleftrightarrow \quad SO(N)_{k} \ {\rm with}\ N_f\ {\rm real\ fermions} \cr
Sp(k+N_f/ 2)_{-N} \ {\rm with}\ N_f\ {\rm scalars} \quad &\longleftrightarrow \quad Sp(N)_{k} \ {\rm with}\ N_f\ {\rm fermions}}}

In all these cases the scalars have quartic interactions.  We flow to the IR and allow to tune the fermion and boson masses to a critical point, if it exists.  Except for some special cases, we cannot prove that such a critical point exists. Many of the statements that we will make about the phases of these three-dimensional theories are independent of whether the corresponding transitions are second order or not.

As we said, the dualities above~\dualities,\dualitiesi\ were conjectured to hold for values of $N_f$ that are bounded from above by, loosely speaking, the Chern-Simons level of the fermionic theory. The central point of this note is to extend this conjecture about the behavior of these theories to larger values of $N_f$. It is logically possible that the previous conjectures are true, but  their newer extended version is not.

Of course such proposals are subject to stringent consistency checks from symmetries, anomalies, mass deformations that allow to decrease $N_f$,
and constraints from the large $N_f$ limits of 2+1 dimensional gauge theories~\AppelquistTC.

Our conjecture is further motivated by the study of domain walls in 3+1-dimensional QCD. This will be developed in great detail in~\PaperTwo. Briefly, pure Yang-Mills theory in 3+1 dimensions at $\theta=\pi$ has a first order transition associated with the spontaneous breaking of time-reversal.  Because of anomaly considerations the domain wall at this transition carries a nontrivial TQFT -- $SU(N)_1$ Chern-Simons theory~\GaiottoYUP. Suppose we now add to the 3+1-dimensional theory heavy quarks with real positive masses, such that time reversal symmetry is preserved by the Lagrangian but broken in the vacuum. For large masses the domain wall is still described by an $SU(N)_1$ Chern-Simons theory, since it is hard to make a topological theory disappear. As we lower the quarks masses, the theory on the domain wall undergos a phase transition. Our scenarios below about the dynamics of the 2+1-dimensional theory are consistent by the transitions on these domain walls~\PaperTwo.

\ifig\SUlargek{The phase diagram of $SU(N)_k$ for $N_f\leq 2k$.  We define the transition point to be at $m=0$.  The gapped phases have topological field theories, $SU(N)_{k\pm {N_f/ 2}}$, which are related by level-rank duality of spin theories to $U(k\pm {N_f/ 2})$ \HsinBLU.  For $k={N_f/ 2}$ the phase for negative $m$ is trivial.
}%
{\epsfxsize4in\epsfbox{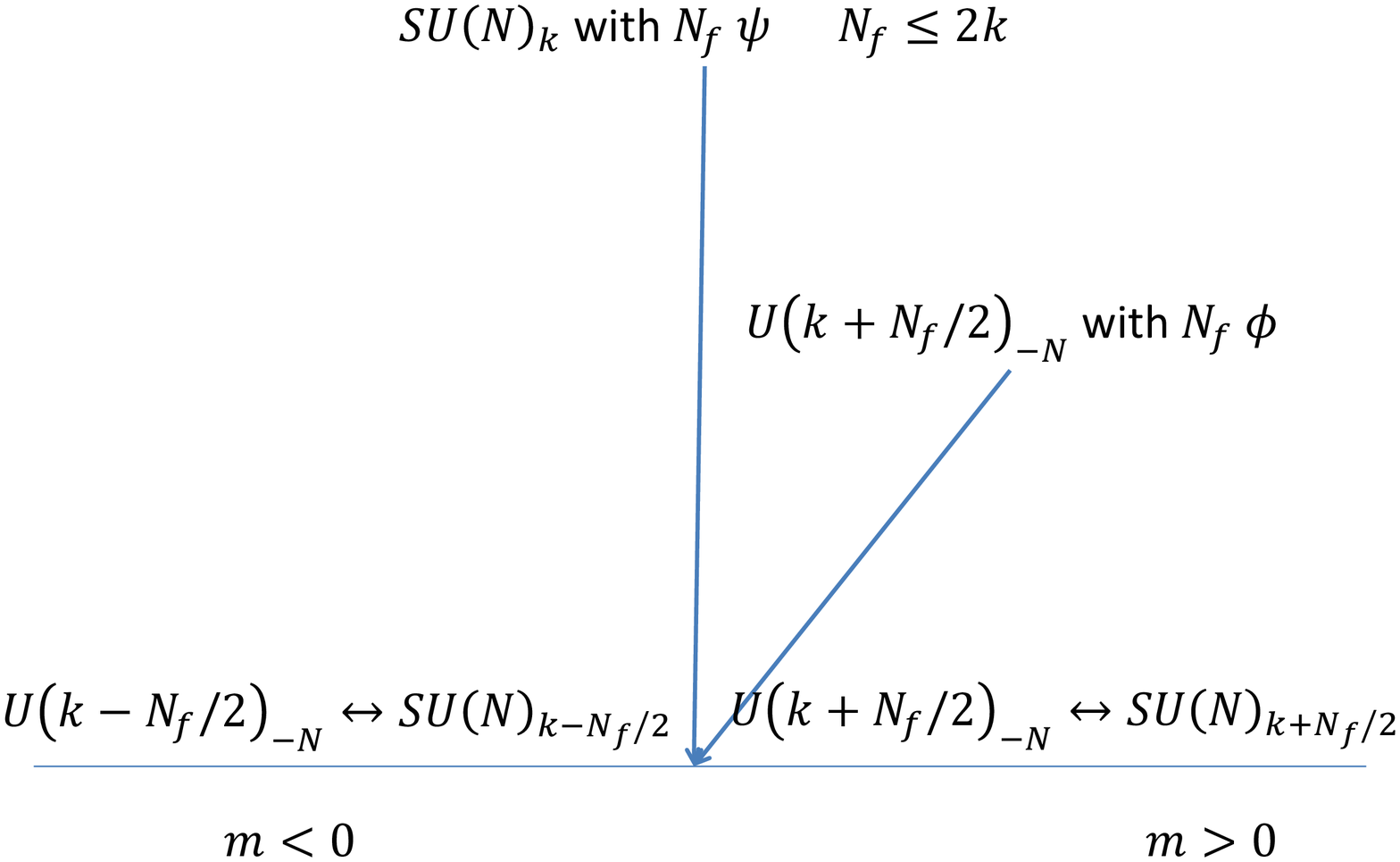}}

\bigskip\centerline{\it Summary of our proposed scenario}
\bigskip

Let us summarize our conjecture for $SU(N)_k+N_f$ fermions, starting from the case $N_f\leq 2k$, which is covered by the dualities~\dualities,\dualitiesi. The theory has two phases, depending on the mass of the quarks. One phase is $SU(N)_{k+N_f/2}$ pure Chern-Simons theory and the other phase is $SU(N)_{k-N_f/2}$ pure Chern-Simons theory. The transition could be first or second order. The global symmetry is never broken. The theory is dual to $U(k+N_f/2)_{-N}+N_f$ bosons, which has exactly these two pure Chern-Simons phases (this can be seen by level-rank duality). For $N_f=2k$ this description still makes sense; on one of the sides of the transition the ground state is trivial. This summarizes the first line in~\dualitiesi. We also summarize this situation in \SUlargek.

For larger $N_f$ this picture breaks down because the bosonic theory has a sigma model at low energies for negative mass squared (for $N_f\leq 2k$ the bosonic model does not have a sigma model phase in its semi-classical regimes and instead it flows to a Chern-Simons TQFT as depicted in \SUlargek). By taking the mass squared of the scalars to be large and negative we can make the sigma model as weakly coupled as we like (i.e.\ the radius of the target space can be as large as we like).
No such weakly coupled sigma model with a large target space exists in the fermionic theory and therefore the duality fails \HsinBLU. Indeed let us turn on equal masses $m$ for all the quarks.  Let us analyze it for $|m|\gg g^2$ (where $g$ is the gauge coupling).  Then we can integrate out the fermions and use \klow\ to find that the low energy theory is gapped and it is described by a pure $SU(N)_{k_{IR}}$ Chern-Simons theory with level
\eqn\largemk{k_{IR}=k+\sign(m)N_f/2~.}
This leaves the interesting question of what the theory does for small values of $m$, of order $g^2$.

\ifig\SUsmallk{The phase diagram of $SU(N)_k$ for $2|k|<N_f<N_*$.  The gapped phases have topological field theories, $SU(N)_{ k\pm {N_f/ 2}}$, which are related by level-rank duality of spin theories to $U( {N_f/ 2}\pm k)_{\mp N}$.  The middle gapless phase is a nonlinear sigma model on a Grassmannian with a Wess-Zumino term $\Gamma$, whose coefficient is $N$. For $N=2$ the Grassmannian is replaced by $Sp(N_f)/( Sp(N_f/2 + k)\times Sp(N_f/2 -k))$.
}%
{\epsfxsize4in\epsfbox{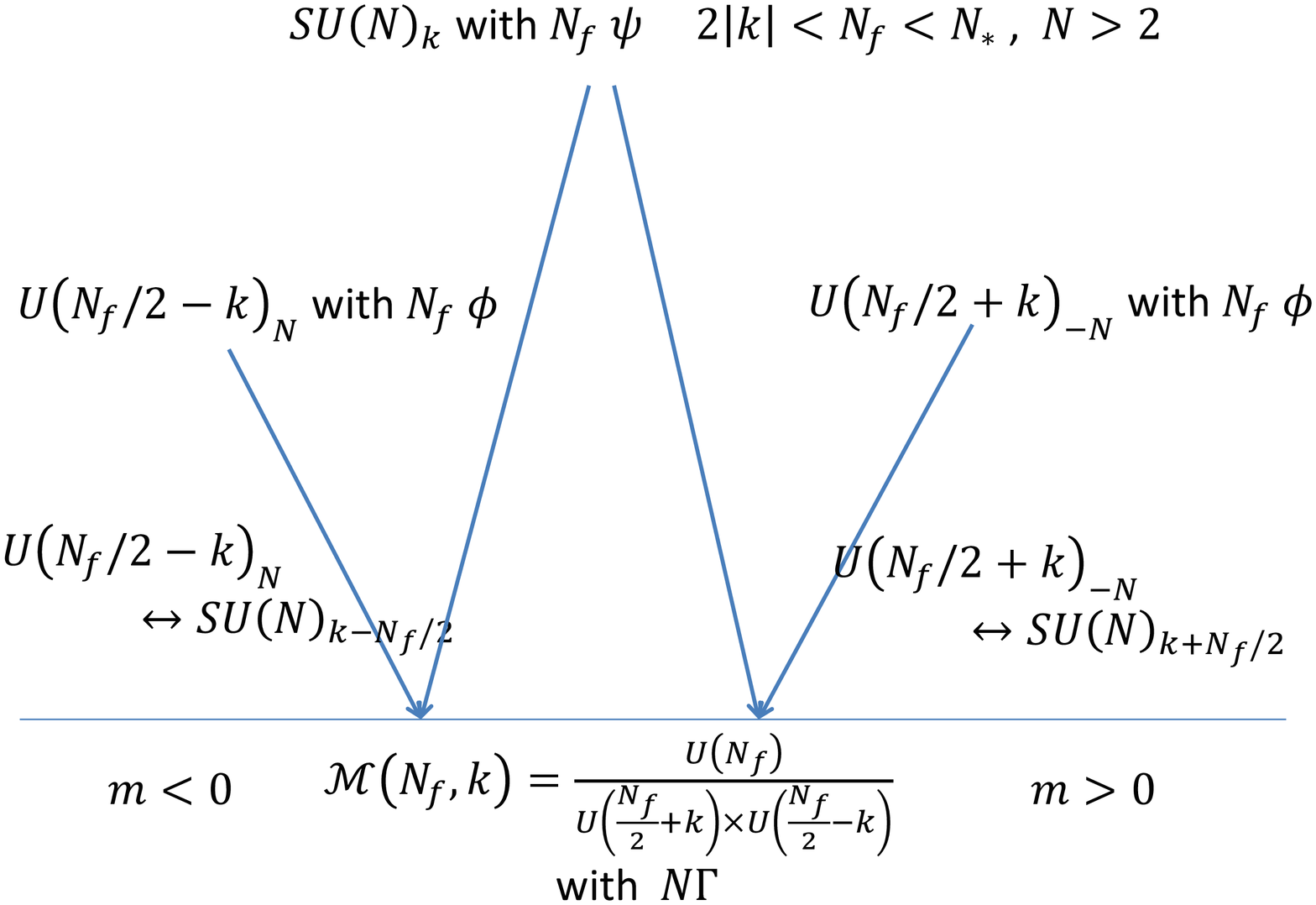}}

Our conjecture is that there exists a window $N_*(N,k)>N_f>2k$ where $N_*$ is an unknown function of the parameters of the theory such that the $U(N_f)$ symmetry that rotates the various flavors\foot{More precisely, only a certain quotient of $U(N_f)$ acts faithfully \BeniniDUS\ (see also~\KomargodskiDMC\ for a similar discussion in a slightly different context) and the system also has charge conjugation symmetry.  These will not play an important role in our discussion.} is spontaneously broken leading to the coset \eqn\GRA{\CM(N_f,k)={U(N_f)\over U(N_f/2+k)\times U(N_f/2-k)}~.}
Our proposal is that this is a purely quantum phase, not visible in the semi-classical limits of the theory and the target space of this sigma model (i.e.\ the complex Grassmannian~\GRA) never becomes large. It is important to mention that we need to add to the sigma model Lagrangian a certain Wess-Zumino term, which we will discuss in detail soon.

In the special case  $k=N_f/2-1$ this result is consistent with the study of domain walls in four dimensions~\PaperTwo. This proposal for $N_*(N,k)>N_f>2k$ {\it implies} the dualities~\dualities,\dualitiesi\ for lower values of $N_f$ and shifted values of $k$, as we explain in the bulk of the paper.

In addition, one can ``derive'' this picture by starting from  $SU(N)_0+N_f$ fermions (with even $N_f$).  It was suggested in~\refs{\VafaTF\VafaXH\VafaXG-\HongSB} that in some range $0<N_f<N_*(N,0)$ this theory breaks its global symmetry leading to a nonlinear sigma model with target space
  \eqn\GRAi{\CM(N_f,k=0)={U(N_f)\over U(N_f/2)\times U(N_f/2)}~.}
The case $k=0$ is therefore an important consistency check of our scenario. But it is not just a consistency check -- we can reverse the logic and assume this scenario at $k=0$ and then derive all the rest.
By deforming the nonlinear sigma model with masses that correspond to fermion masses in the ultraviolet theory, one can see that this breaking must take place in a whole region of parameter space (as we vary the symmetry-preserving mass deformation) and not just at one point, and, in addition, one can derive all the other symmetry breaking patterns~\GRA. In particular, as we will explain, if the symmetry breaking~\GRAi\ does not occur, then our conjecture is necessarily wrong.

In addition, we stress again that we need to add to this discussion the Wess-Zumino term. There are various ways to see that this Wess-Zumino term is necessary, for example, by considering the quantum numbers of Skyrmions (baryons). This derivation also leads to interesting constraints on $N_*(N,k)$, which we discuss in section 2.

The sigma model~\GRA\ arises from a condensation of the quark bilinear. Indeed, consider the condensate
\eqn\psipsiv{\psi\psi^\dagger=diag(x,...,x,y,...,y)~,}
with $x$ appearing $N_f/2+k$ times and $y$ appearing $N_f/2-k$ times ($y\neq x$). It breaks the symmetry to $U(N_f/2+k)\times U(N_f/2-k)$. Therefore, we obtain the coset~\GRA. In the interesting special case of $k=0$ the model has time reversal symmetry at $m=0$. The symmetry breaking phase~\GRAi, if it exists for the given value of $N_f$, must contain the point where the quarks are massless.

The fermionic theory therefore has two special points (see \SUsmallk), one where it makes a transition from the semi-classical phase $SU(N)_{k+N_f/2}$ to the phase~\GRA\ and the other is where it makes a transition from~\GRA\ to the second semi-classical phase, $SU(N)_{k-N_f/2}$. We propose that the former is dual to scalar theory $U(N_f/2+k)_{-N}+N_f$ scalars and the latter to the theory $U(N_f/2-k)_N+N_f$ scalars. As a simple consistency check, note that both of these scalar theories lead to the coset~\GRA.
In other words, the fermionic theory covers the whole parameter space, while each of its bosonic duals describes only a patch of the parameter space -- the patch around each transition.  This is consistent with the fact that all three theories have the same global symmetry and the same 't Hooft anomalies~\BeniniDUS. Again, we do not have a clear picture of whether these phase transitions are first or second order, but the symmetry breaking phase should be robust and independent of the question of the order of the transition.

This  proposal for the dynamics of QCD in 2+1 dimensions for
$N_*(N,k)> N_f>2k$ is summarized in \SUsmallk. The phases of the theory that are described in the deep infrared by pure Chern-Simons theory are intuitively not confining as the low energy observer can see the Wilson lines of quarks. The symmetry breaking phase at small $|m|$ should be viewed as a confining phase. Therefore, confinement takes place for $N_*(N,k)> N_f>2k$, but it does not take place for $N_f\leq 2k$.\foot{There is no rigorous order parameter for confinement here, so this discussion is only at the intuitive level. In the case of $SO(N)$ gauge theory (more precisely, $Spin(N)$) there is a rigorous notion of confinement and what we said in the text is rigorously true for the analogous phases there.  See section 3.}

Finally, it remains to discuss what happens for $N_f\ge N_*$. We know that for sufficiently large $N_f$ the theory has to be conformal, with only Chern-Simons phases (as in \SUlargek) on both sides of the transition~\AppelquistTC.  We therefore assume that this is the behavior of the theory for all $N_f\ge N_*$. No scalar dual is known in this regime. $N_*$ is therefore the critical number below which (and above $2k$) the symmetry breaks in an intermediate, quantum phase, that is invisible semi-classically.

Our proposal for the phases of $SO$ and $Sp$ gauge theories is rather similar in spirit, with an intermediate symmetry breaking phase for sufficiently large (but not too large) values of $N_f$.

In section 2 we discuss some additional details about $SU(N)_k$ with $N_f$ fermions in the fundamental representation. In section 3 we discuss the dynamics and phases of the $SO$ and $Sp$ theories, where there are a few novelties.

\newsec{More on $SU(N)_k$ with $N_f$ Fermion Flavors}

In this section we discuss a number of topics related to the proposed Grassmannian phase of $SU(N)_k$ with $N_f$ fermion that have not been described in detail in the introduction.

\subsec{Studying the sigma model}

We will study some properties of the sigma model with target space
 \eqn\GRASM{\CM(N_f,k)={U(N_f)\over U(N_f/2+k)\times U(N_f/2-k)}~.}
One way to think about it is intrinsic to this sigma model and uses properties of this space.  This approach is standard in the literature. Alternatively, we find it convenient to view this theory as the low energy approximation of a gauged linear sigma model.  In the special case of $\CM(N_f,k=N_f/2-1) =\C\P^{N_f-1}$ this approach is also standard in the literature.

Specifically, we can think about~\GRASM\ as the low energy theory of the gauge theory
\eqn\GLSM{U(N_f/2-k)+N_f \ {\rm scalars}\ \Phi^i~,}
where we add large negative mass squared for the scalars, such that the model is weakly coupled and we can treat it classically.
Of course, as we explained in the introduction, this model, when appropriately modified by Chern-Simons terms, plays a more fundamental role in the story than just an auxiliary linear sigma model. However, here we will use it to study some properties of the Grassmannian per se without referring to the more general role this model plays (upon modifying it with Chern-Simons terms and allowing it to be strongly coupled).  Later we will add the Chern-Simons term and show that it leads to a Wess-Zumino term in the non-linear model.

Following the standard treatment in the $\C\P^{N_f-1}$ model, it is straightforward to write down explicitly the two-derivative nonlinear sigma model Lagrangian with target space~\GRASM. The fundamental degree of freedom is a bi-fundamental $U(N_f/2+k)\times U(N_f/2-k)$ scalar $\pi$ and the Lagrangian is \eqn\twoder{{\cal L}_{kinetic}\sim  Tr\biggl(  \del \pi\del\pi^\dagger (1+\pi\pi^\dagger)^{-1} -\del \pi \pi^\dagger (1+\pi\pi^\dagger)^{-1}\pi\del\pi^\dagger (1+\pi\pi^\dagger)^{-1}  \biggr)~.}
This metric can be derived from the K\"ahler potential
\eqn\Kahlergr{K = Tr\log(1+\pi\pi^\dagger)~.}

Instead, we will continue to analyze the gauged linear sigma model with large, degenerate negative mass squared for the scalars, such that the model flows to~\GRASM.

\bigskip
\centerline{\it Wess-Zumino Terms and Skyrmions}
\bigskip

Here we describe the Wess-Zumino terms that must be added to the nonlinear sigma model Lagrangians.

We start from the simplest Grassmannians, namely the $\C\P^{N_f-1}$ manifolds, and then briefly describe the more general story.  As above, we describe it in terms of a $U(1)$ gauge theory with a gauge field $b$ with $N_f$ scalars.  In this theory we can add the Chern-Simons terms
\eqn\CSt{\eqalign{
&{N\over 4\pi} bdb +{1\over 2\pi} db B \qquad {\rm for} \qquad N\in 2\Z \cr
&{N\over 4\pi} bdb +{1\over 2\pi} db A \qquad {\rm for} \qquad N\in 2\Z +1~,\cr
}}
where $A$ is a classical background spin$_c$ connection and $B$ is a classical background $U(1)$ gauge field.  For even $N$ each term in \CSt\ is separately meaningful.  The same is true for odd $N$ on a spin manifold.  But on a non-spin manifold with odd $N$ only the sum of the two terms in \CSt\ is meaningful.  (See \refs{\SeibergRSG-\HsinBLU} for more details.)
The equations of motion of $b$ set
\eqn\beom{db = w  + \dots }
where the ellipses denote higher order terms\foot{The low energy nonlinear model is characterized by a scale that originates from the expectation value of the scalars in the linear model.  It appears in front of the kinetic term in the nonlinear model.  These higher order terms are suppressed by inverse powers of that scale.} and $w$ is the pull back of the K\"ahler form of the $\C\P^{N_f-1}$ manifold to spacetime, normalized such that its integrals over two-cycles $\Sigma_2$ are
\eqn\wnorm{\int_{\Sigma_2} w\in 2\pi\Z~.}
For $N_f>2$ we can substitute \beom\ back in the Lagrangian and find
\eqn\CSte{\eqalign{
&{N\over 4\pi}\int_{M_4}ww +{1\over 2\pi} \int_{M_3}w B \qquad {\rm for} \qquad N\in 2\Z \cr
&{N\over 4\pi}\int_{M_4} ww +{1\over 2\pi}\int_{M_3} w A \qquad {\rm for} \qquad N\in 2\Z +1~,\cr
}}
where $M_3$ is our spacetime and $M_4$ is a four manifold whose boundary is $M_3$.  The first term can be interpreted as a Wess-Zumino term in the nonlinear model. We see that on a non-spin manifold $N$ should be even.  If $N$ is odd, we need to have a spin$_c$ structure with a connection $A$. In terms of the massless pions the Wess-Zumino term induces interactions such as
\eqn\WZc{\sim N \int d^3 x \epsilon^{\mu\nu\rho} (\del_\mu\pi^\dagger\cdot\pi)(\del_\nu\pi^\dagger\cdot\del_\rho\pi)+\cdots~,}
where the ellipses stand for higher order interaction terms (with three derivatives) that are necessary for $SU(N_f)$ invariance.

The background fields $A$ and $B$ couple in \CSt\ to monopole operators of the linear model.  The spin of the monopole is $N/2$ and hence, for even $N$ it is a boson, which couples to an ordinary background gauge field $B$ and for odd $N$ it is a fermion, which couples to a spin$_c$ connection $A$.  In the nonlinear model these background fields couple to configurations with nonzero $\int_{\Sigma_2} w$.  If $\Sigma_2$ is our space, then these charged objects are Skyrmions.  Therefore, the Skyrmions of the model are bosons or fermions depending whether $N$ is even or odd.

Let us consider the baryons of the fermionic $SU(N)$ theory. They are created by acting with the baryon operator on the vacuum. In terms of the bosonic dual, we can act on the vacuum with the (appropriately dressed) monopole operator. Since the monopole operator is dual to the baryon operator, we get dual excitations in the Hilbert space~\refs{\RadicevicYLA,\AharonyMJS}.
In our case we can further observe that the monopole operators flows to the Skyrmion operator in the sigma model phase. Therefore, the baryon particles can be identified with the Skyrmions of the non-linear sigma model.\foot{See a related discussion in~\HongSB.}
As a check, the quantum numbers of these baryon operators match with those of the monopole operators of the bosonic theory and with the Skyrmion operators of the low energy theory.\foot{The $SU(N_f)$ classical gauge fields already couple to the two-derivative Lagrangian, which is the reason we do not discuss them here. By contrast, the baryon symmetry does not act on  the two-derivative Lagrangian. For completeness it is however worth mentioning that the $SU(N_f)$ gauge fields do couple to Skyrmions, since the baryon carries an $SU(N_f)$ representation that is necessarily nontrivial, i.e.\ with nonzero $N_f$-ality, if $gcd(N_f,N_c)\neq 1$.}  This identification of the Skyrmions with the microscopic baryons and the way their quantum numbers are determined by $N$ through the Wess-Zumino term are exactly as in the similar four-dimensional theory \refs{\WittenTW,\WittenTX}. In addition, as in four dimensions, the Wess-Zumino term breaks a symmetry that exists accidentally in the nonlinear-sigma model. For $k\neq 0$ it is time-reversal symmetry and for $k=0$ it is another discrete symmetry (exchanging the two $U(N_f/2)$ factors, which acts as transposition on the matrix of pions).

A very important exception occurs when $N_f=2$. Clearly, the term $ww$ in~\CSte\ vanishes then.  There is however still a discrete Wess-Zumino action, which can be further extended to non-spin manifolds by allowing $A$ to be a spin$_c$ connection. The construction of this discrete term will be discussed in~\PaperThree. Note that in~\WilczekCY\ a non-local modification of the $\C\P^1$ action was described and it was shown to affect the quantum numbers of the Skyrmions. The relation of this to the discrete invariant of~\PaperThree\ will be described there.

Now let us discuss briefly the general case of $\CM(N_f,k)={U(N_f)\over U(N_f/2+k)\times U(N_f/2-k)}$, which arises in $SU(N)_k$ gauge theory with $N_f$ fermions. The main novelty of the general case compared to  $\C\P^{N_f-1}$ is that the $H^4$ cohomology\foot{For a more precise treatment of this Wess-Zumino term see~\FreedMX.}
 is generated by two four-forms and not just one. Therefore, there are a priori two Wess-Zumino terms. It is easiest to describe these two coefficients using the linear sigma model description, where they are the standard $U(N_f/2-k)$ invariants, $\int_{M_4} Tr F\wedge Tr F$ and $\int_{M_4} Tr F\wedge F$. In our proposal we take only the latter one with coefficient proportional to $N$.

Let us finish this discussion with a comment about strings in the sigma model. Such strings exist, if the fundamental group of the target space is non-trivial.  Then consider a configuration that is independent of one direction, say $x^1$, and wraps the nontrivial cycle in $x^2$.  The energy density is localized in a small region in $x^2$ and therefore such a configuration is a strings in $\R^2$.  This fundamental group and the corresponding strings are associated with an unbroken one-form global symmetry~\refs{\KapustinGUA,\GaiottoKFA}. In the $SU(N)_k+N_f$ fermions theory there is no such one-form symmetry at short distances  and it is therefore nice to find out that indeed $\pi_1(\CM(N_f,k))$ vanishes.

\subsec{Flowing down in $N_f$}

We now study some mass perturbations of our model. Instead of perturbing the nonlinear model we will perturb the gauged linear model by a small mass term $\delta m^2_{i\bar j}\Phi^i\bar \Phi^{\bar j}$.

One simple perturbation $\delta m^2_{i\bar j} \sim \delta_{i\bar j}$ preserves the $SU(N_f)$ global symmetry.  It just shifts the mass squared of the scalars, and in the semi-classical regime they would still condense and lead to the Grassmannian~\GRASM, albeit with a different size. The Grassmannian should therefore remain when a mass term proportional to the identity matrix is added to the Lagrangian. This fact has some ramifications for the mapping of parameters between the Grassmannian Lagrangian, the gauged linear sigma model, and the fermionic theory. In essence, that means that we can mix general mass perturbations, which are in the adjoint representation, with the singlet representation when we map the parameters.

Next, we analyze two mass perturbations that decrease $N_f$ to $N_f-1$.
First, consider a mass deformation with the only nonzero entry being $\delta m^2_{1\bar 1}>0$. Then the $N_f-1$ scalars with degenerate masses would still condense. The scalar $\Phi^1$ is now massive around this configuration. Therefore we obtain the coset
\eqn\posmass{\CM(N_f-1,k-1/2)={U(N_f-1)\over U(N_f/2+k-1)\times U(N_f/2-k)}~.}

Second, let $\delta m^2_{1\bar 1}<0$ and small. Then we condense first $\Phi^1$ and obtain a $U(N_f/2-k-1)$ gauge theory coupled to $N_f-1$ scalars. Then, condensing them we obtain the coset
\eqn\negmass{\CM(N_f-1,k+1/2)={U(N_f-1)\over U(N_f/2+k)\times U(N_f/2-k-1)}~.}
Clearly, if we started with $N_f/2-k=1$ this resulting theory is trivial.

Here we have treated the linear sigma model as a proxy for the nonlinear sigma model Lagrangian with target space~\GRASM.  Clearly, instead of analyzing these deformations in the linear model, we could have analyzed them directly in the nonlinear model.  That would have led directly to \posmass\negmass.

Returning to our general proposal, the two mass deformations
\eqn\massdeform{(N,N_f,k)\to (N,N_f-1,k\pm 1/2)}
show that if our proposal about the phase diagram is right for $(N,N_f,k)$, it must also be right for $(N,N_f-1,k\pm 1/2)$.  This fact has two implications that we discuss now.

\bigskip
\centerline{\it Constraints on $N_*(N,k)$}
\bigskip

We can use~\massdeform\ in order to obtain interesting constraints on $N_*(N,k)$.

First, using the same argument as in \AppelquistTC, it is clear that for large enough $N_f$ with or without large $k$ the system has only two phases separated by a second order point.  For $N_f\le 2k$ in this range we have a proposed bosonic dual description of this phase transition.

Next, we consider finite $N_f$ and look for the Grassmannian phase with $2k<N_f<N_*$.  Assuming we found such a point, requiring that the deformations of the Grassmannian theory by a small mass always agree (as above) in the infrared with deformation of the full theory by a large mass parameter, we get a nontrivial constraint on $N_*(N,k)$.
We see that if $N_f<N_*(N,k)$, then it must also be true that $N_f-1<N_*(N,k\pm\half)$, and therefore, for all $N,k$ (with small enough $k$)
\eqn\cons{N_*\left(N,k\right)-1\leq N_*\left(N,k\pm\half\right)~.}

Interestingly, the constraint~\cons\ implies that the size of the window for the Grassmannian phase, namely, $N_*-2k$, is necessarily maximized at $k=0$ (i.e.\ the size of the window can only decrease as $k$ is increased).  In particular, if there is no symmetry breaking at $k=0$ according to $U(N_f/2)\to U(N_f/2)\times U(N_f/2)$, then there cannot be symmetry breaking of the sort that we discuss at any $k$. Furthermore, $N_*$ cannot decrease too fast (since $N_*\left(N,k+1/2\right)\leq N_*\left(N,k\right)+1$, the average derivative with which it can decrease is not bigger than 2 in absolute value).

\bigskip
\centerline{\it Going up in $k$}
\bigskip

Next, we use \massdeform\ to offer an alternative point of view on our proposal. Let us start with the $SU(N)_0+N_f$ fermions theory with an even number of fermions.  This theory is time-reversal invariant for $m=0$.  It has been suggested that its global $U(N_f)$ symmetry is spontaneously broken to $U(N_f/2)\times U(N_f/2)$.

One argument for it is that we can give the fermions a time-reversal invariant mass preserving this unbroken symmetry.  We do that by giving $N_f/2$ of them mass $+m$ and $N_f/2$ of them mass $-m$.  This way we can gap the system in a smooth way.  Such a deformation was used in  \refs{\VafaTF,\VafaXH,\VafaXG} to argue that time reversal and the $U(N_f/2)\times U(N_f/2)$ symmetry should remain unbroken.  Of course, his does not mean that $U(N_f)$ must be broken.

A more abstract point uses the fact that this theory is time-reversal invariant as well as $U(N_f)$ invariant. However, there is a mixed 't Hooft anomaly between these two symmetries. This is just a variant of the standard mixed anomaly between time reversal and fermion number symmetry~\RedlichDV.  For $N_f\geq N_*$ this anomaly is represented in the IR theory by a nontrivial conformal field theory. At $N_f<N_*$ the anomaly can be saturated by breaking $U(N_f) \to U(N_f/2)\times U(N_f/2)$.   It does not mean that the range $N_f<N_*$ exists\foot{In principle, there could be other ways to match the anomaly in the infrared, e.g.\ with a nontrivial TQFT and an invariant vacuum or with a nontrivial CFT.} (i.e.\ it could be that $N_*=0$), but let us assume it does.
Hence, this symmetry breaking pattern must exist in a range of the $U(N_f)$-preserving mass parameter, and not just at $m=0$ as in four dimensions.

Next, we can deform the model as in~\massdeform\ and arrive at the other Grassmannians with other values of $N_f$ and $k$.  More precisely, this way we can get only to $N_f<N_*(N,0)-2k$. We conclude that assuming the $U(N_f) \to U(N_f/2)\times U(N_f/2)$ breaking in the $k=0$ theory, we derive our more general proposal about the intermediate phase. A central nontrivial point is that already the $k=0$ model needs to be supplemented with the Wess-Zumino term we described above. Then, its existence for the other values of $k$ follows.

\newsec{$SO(N)_k$ and $Sp(N)_k$ with $N_f$ Fermion Flavors}

The situation for $SO(N)$ and $Sp(N)$ gauge theories is similar to the $SU(N)$ situation described above.  Here we use
\eqn\SOSpdualities{\eqalign{
SO(N)_{k} \ {\rm with}\ N_f\ {\rm fermions} \quad &\longleftrightarrow \quad SO(k+{N_f\over 2})_{-N} \ {\rm with}\ N_f\ {\rm scalars}\cr
SO(N)_{k} \ {\rm with}\ N_f\ {\rm fermions} \quad &\longleftrightarrow \quad SO(-k+{N_f\over 2})_N \ {\rm with}\ N_f\ {\rm scalars} \cr
Sp(N)_{k} \ {\rm with}\ N_f\ {\rm fermions} \quad &\longleftrightarrow \quad Sp(k+{N_f\over 2})_{-N} \ {\rm with}\ N_f\ {\rm scalars}\cr
Sp(N)_{k} \ {\rm with}\ N_f\ {\rm fermions} \quad &\longleftrightarrow \quad Sp(-k+{N_f\over 2})_N \ {\rm with}\ N_f\ {\rm scalars}}}

\ifig\SOlargek{The phase diagram of $SO(N)_k$ for  $N_f\le 2k$.
}%
{\epsfxsize4in\epsfbox{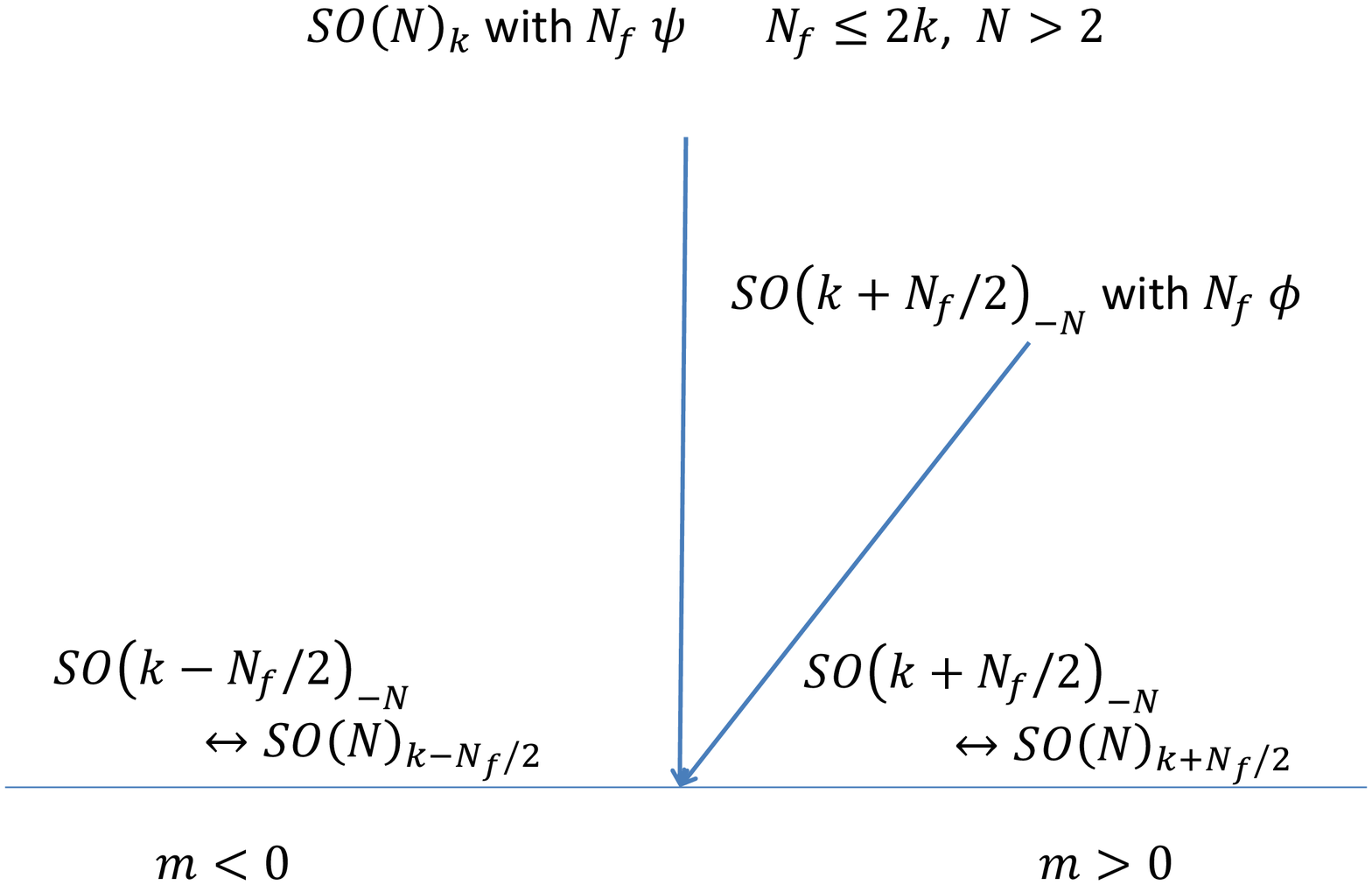}}

\ifig\SOsmallk{The phase diagram of $SO(N)_k$ for $2|k|<N_f$.
}%
{\epsfxsize4in\epsfbox{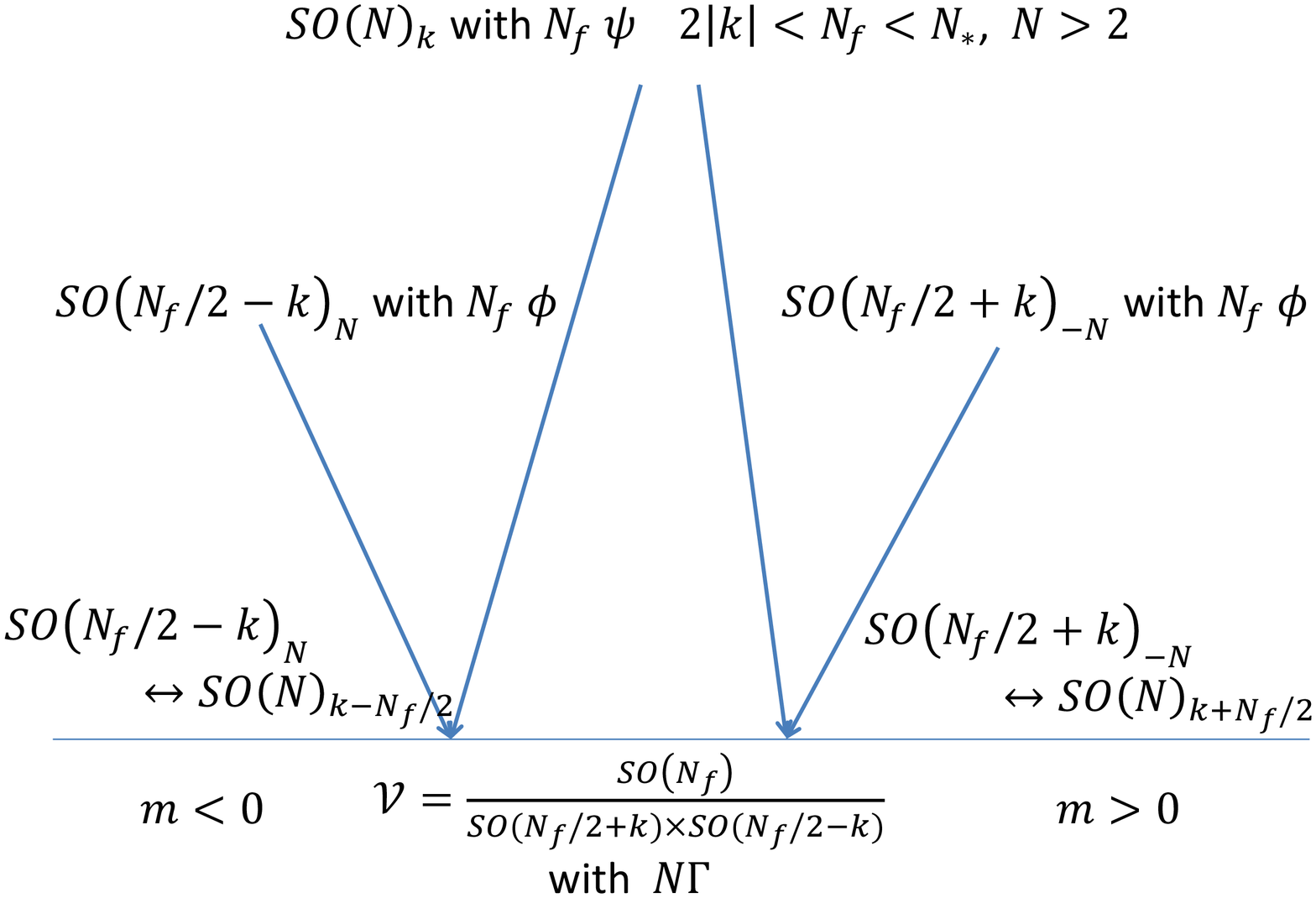}}

\ifig\Splargek{The phase diagram of $Sp(N)_k$ for $N_f\le 2k$.
}%
{\epsfxsize4in\epsfbox{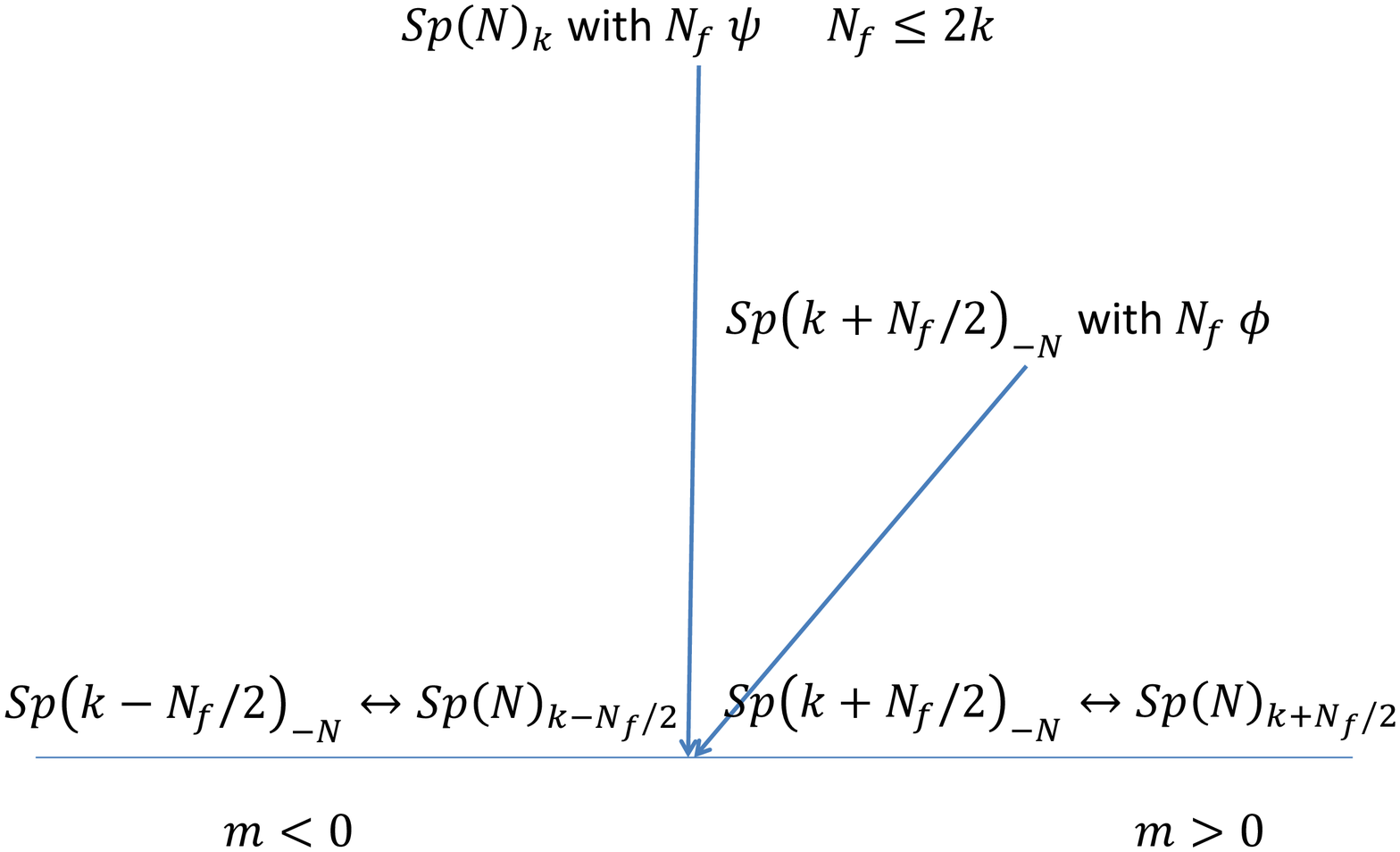}}

\ifig\Spsmallk{The phase diagram of $Sp(N)_k$ for $2|k|< N_f<N_*$.
}%
{\epsfxsize4in\epsfbox{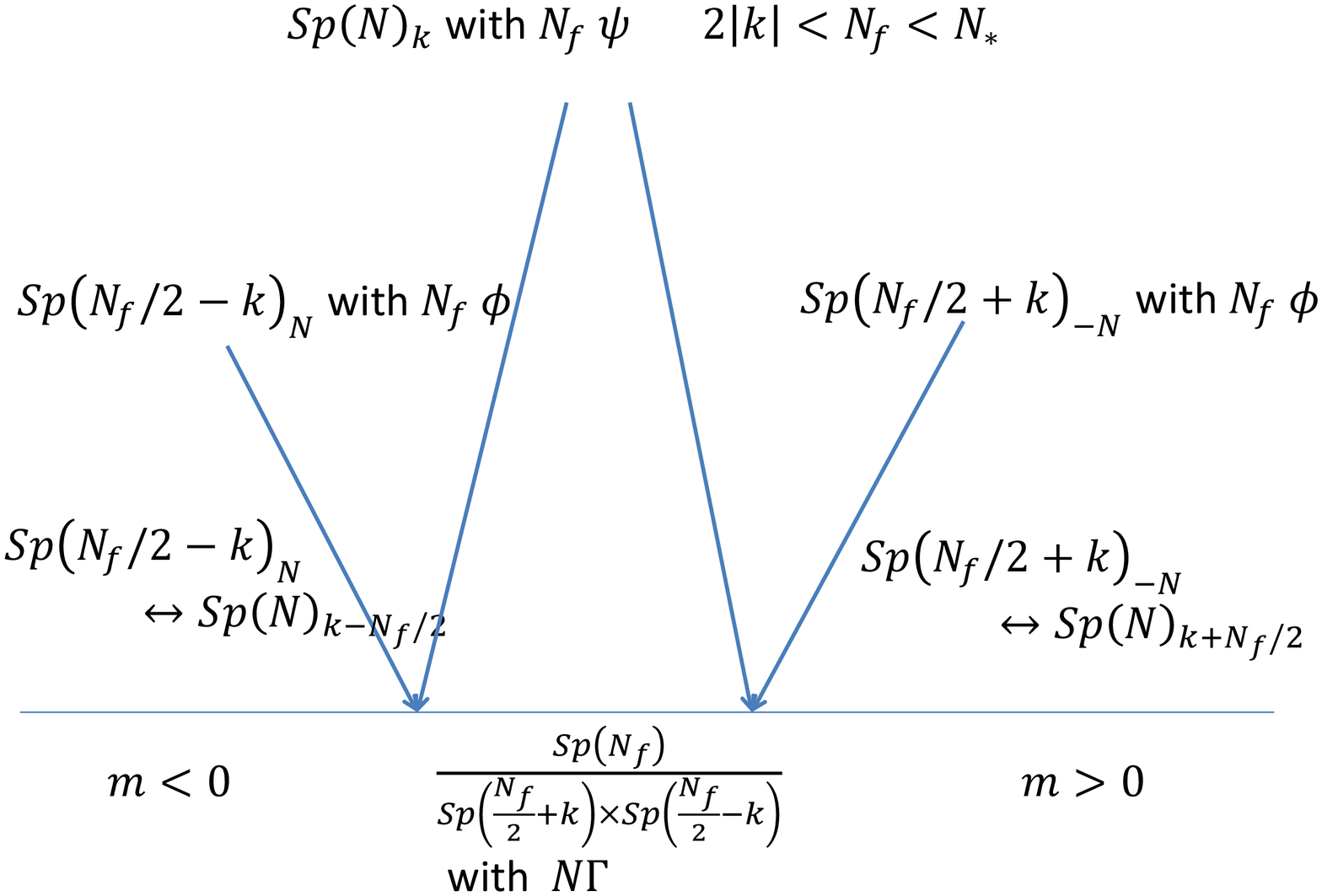}}

Again, we expect some $N_*(N,k)$ such that for $N_f\ge N_*$ there are only two non-confining phases with a second order point between them.\foot{The value of $N_*(N,k)$ can depend on the gauge group.  We use the same notation as in the $SU(N)$ discussion above and suppress the dependence on the gauge group.} For $2k<N_f<N_*$ we expect three phases with a gapless middle phase describing the spontaneous breaking of the global symmetry.  And for $N_f\le 2k$ the intermediate phase is absent.  For $SO(N)$ it is depicted in \SOlargek\ and \SOsmallk, and for $Sp(N)$ it is depicted in \Splargek\ and \Spsmallk.

All the comments we made in the previous section still apply, but there are a few additional noteworthy facts.  The $SO(N)$ fermionic theory has a $\Z_2$ magnetic global 0-form symmetry.  Under the duality it is mapped to a charge conjugation symmetry -- the symmetry that extends the $SO({N_f\over 2}\pm k)_N$ gauge symmetry to $O({N_f\over 2}\pm k)$ \AharonyJVV.  In the topological phases with $SO(N)_{k\pm N_f/2}$ this symmetry is unbroken.

However, in the Grassmannian phase the magnetic $\Z_2$ symmetry is spontaneously broken.  This can be seen most easily using the scalar dual theories, where the Grassmannian can be seen at tree level.   Without loss of generality consider the scalar theory with gauge group $SO({N_f\over 2}- k)$. In the Grassmannian phase the scalars have an expectation value, which, up to symmetry transformations, is of the form $\phi_{ia} = x \delta_{ia}$ where $i$ labels the colors and $a$ labels the flavors, and $x$ is nonzero. This breaks the $O(N_f)$ global symmetry to $SO({N_f\over 2}- k) \times O({N_f\over 2}+k)$, where the first factor is a diagonal subgroup between a subgroup of the global symmetry group and the gauge group.  This leads to the vacuum manifold
\eqn\CVde{\CV ={SO(N_f)\over SO({N_f\over 2}+k)\times SO({N_f\over 2}-k)}~,}
as in \SOsmallk.  The point $\phi_{ia} = x \delta_{ia}$ and the point where the first entry, $\phi_{11}$, flips sign are related by a broken global symmetry transformation reflecting the spontaneously broken charge conjugation symmetry of the bosonic theory.

 Alternatively, we can parameterize the vacuum manifold by a gauge invariant order parameter of the form $\phi\phi$.  It lives in
\eqn\CVtde{\tilde \CV={SO(N_f)\over S[O({N_f\over 2}+k)\times O({N_f\over 2}-k)]}~.}
The true target space $\CV$ is a double cover of this space because every point in $\tilde \CV$ corresponds to two different points in $\CV$, which differ by the expectation value of a gauge invariant baryon operator of the form $\phi\phi...\phi$, where the color indices are contracted using an epsilon symbol.  This order parameter is odd under the charge conjugation symmetry, thus establishing that this symmetry is spontaneously broken.  Using the duality it is mapped to a monopole operator of the underlying $SO(N)$ gauge theory.  Its expectation value shows that the magnetic symmetry of the fermionic $SU(N)$ theory is spontaneously broken.

We see that the magnetic symmetry of the fermionic theory is unbroken in the topological phase and it is broken in the Grassmannian phase.  This is consistent with the intuitive picture that the topological phases are not confining, but the Grassmannian phase is confining (we typically think of a phase with broken magnetic symmetry as confining).  A more precise way to see that uses one-form global symmetries \refs{\KapustinGUA,\GaiottoKFA}.  None of our theories have such a symmetry.  This follows from the fact that the gauge group acts faithfully on the matter fields. In particular, the space~\CVde\ is simply connected, which is encouraging, since that means that there are no stable strings in the macroscopic theory.  However, we can change the $SO(N)$ theories and replace them by $Spin(N)$ theories (by gauging the discrete 0-form $\Z_2$ magnetic symmetry, as described in the introduction) and then these theories have a ``bonus'' $\Z_2$ one-form global symmetry.  The charged objects under this one-form symmetry are Wilson lines in the spinor representation of $Spin(N)$.  They can be used to diagnose confinement rigorously.

In order to change the gauge group to $Spin(N)$ we turn the global $\Z_2$ magnetic symmetry of the $SO(N)$ theories to a gauge symmetry.   Then the topological phases are $Spin(N)_{k\pm N/2}$ and the existence of nontrivial spinor Wilson lines in these phases confirms our assertion that these phases are not confining.

The Grassmannian phase is more interesting.  Again, we analyze it using the two dual scalar theories.  Here the $\Z_2$ magnetic symmetry of the fermionic theory acts as charge conjugation \AharonyJVV\ and correspondingly the gauge group is $O(N_f/2 \pm k)$.\foot{For a more detailed discussion see \CordovaVAB.}   The effect of this extension of the gauge group on the Grassmannian is to mod it out by $\Z_2$ turning it from $\CV$ of \CVde\ to $\tilde \CV=\CV/\Z_2$ of \CVtde.  The resulting space is not simply connected $\pi_1(\tilde \CV)=\Z_2$.\foot{ The fact that the moduli space of vacua in the $SO(N)$, $Spin(N)$, and $O(N)$ theories are related by certain quotients and its relation to duality has already been noted in supersymmetric theories in \refs{\AharonyHDA,\AharonyKMA}.}

We see that the underlying $\Z_2$ one-from global symmetry, which is associated with the center of $Spin(N)$, is realized as a one-form global symmetry in the macroscopic theory.  Because of this nontrivial $\pi_1$ the macroscopic theory has stable $\Z_2$ strings.  These are configurations of the scalars that are independent of one spatial direction and the other spatial direction winds around the non-contractible cycle in the target space.  These charged strings are interpreted as the $\Z_2$ confining strings of the microscopic theory.  We see that in the Grassmannian phase the one-form global symmetry is unbroken and as described in \GaiottoKFA, this means that the theory is confining. A similar comment in a closely related four-dimensional context was made in \refs{\WittenTW,\WittenTX}.

In general, when we gauge an electric one-form symmetry we obtain in the new gauge theory a bonus magnetic zero-form symmetry.  The reverse process was described in the introduction. If the electric one-form symmetry is unbroken, the theory confines.  In this case the bonus zero-form magnetic symmetry of the new theory is expected to be broken.  Similarly, if the electric one-form symmetry is broken and hence the theory does not confine, the bonus magnetic symmetry of the new theory is unbroken. This is the standard relation between spontaneously broken magnetic symmetry and confinement.

The original $SO(N)$ gauge theory (or the $Spin(N)$ theory we have just discussed) has a $\Z_2$ charge conjugation symmetry that extends $SO(N)$ to $O(N)$.  It is mapped under the duality to the $\Z_2$ magnetic symmetry of the bosonic gauge theory \AharonyJVV.  This symmetry is realized in the macroscopic Grassmannian phase as associated with a non-trivial $\pi_2$ of the target space.  The charged operators under this symmetry are baryon operators in the fermionic gauge theory.  They are mapped to monopole operators in the bosonic gauge theory.  And in the low energy Grassmannian sigma model they are mapped to operators creating nontrivial $\pi_2$.  This $\pi_2$ of the Grassmannian leads to Skyrmions, which are odd under this $\Z_2$ symmetry.  They are identified with the baryons of the underlying fermionic gauge theory.

An interesting exception occurs when one or both of the factors in the denominator of \CVde\ is $SO(2)$.  In that case $\pi_2(\CV)$  has one or more factors of $\Z$.  For example, for $N_f=4$, $k=0$ we have $\pi_2\left(SO(4)\over SO(2)\times SO(2)\right)=\Z\times \Z$.  In this case the Grassmannian theory should be supplemented with Skyrmion operators that allow Skyrmions to decay, making the global Skyrmion number only $\Z_2$.  In the case where the bosonic dual has a $SO(2)=U(1)$ gauge symmetry, this enhanced  $\pi_2$ is associated with an enhanced $U(1)_T$ magnetic symmetry.  In this case this symmetry should be explicitly broken to $\Z_2$  by adding monopole operators of charge two to the Lagrangian in order to preserve the duality (as was done recently in other models in \refs{\KomargodskiDMC,\KomargodskiSMK}). These operators have a negligible effect in the Higgs phase, except that they allow Skyrmions to decay, while preserving the remaining $\Z_2$ charge. However, these operators can become important, if their dimension becomes relevant in the IR.

In summary, in the generic case $\pi_2(\CV)=\Z_2$ is nicely consistent with the expectations from the fermionic theory. In the special cases where $N_f/2-k=2$ or $N_f/2+k=2$ we have to add to the bosonic description even monopole operators. Analogously, the nonlinear sigma model Lagrangian needs to be modified by adding monopole-like operators, which would allow even Skyrmions to decay to the vacuum.

Finally, as in the $SU(N)$ theories, also here the Grassmannian~\CVde\ has to be accompanied by certain Wess-Zumino terms that are responsible for the quantum numbers of Skyrmions. We do not describe them in detail here, except to note that they follow from the Chern-Simons terms in the bosonic gauged linear models.

\bigskip
\centerline{\it A Comment about $SO(2)$}
\bigskip

In the discussion of~\SOsmallk\ we restricted ourselves to $N>2$. In fact, our results nicely apply to $N=2$ when properly interpreted. Here we would like to explain how this comes about.  As we will see, not only is our general picture correct also for $N=2$, this case is actually on stronger footing than the more general case and thus supports our general picture.

First, let us explain why we restricted our discussion to $N>2$. In the case of $SO(2)\cong U(1)$ gauge theory with $N_f$ flavors, the symmetry of the massless theory is locally $SU(N_f)\times U(1)$, where the first factor is the flavor symmetry and the second is the magnetic symmetry. This differs from the $SO(N)$ series with $N>2$, where the magnetic symmetry is discrete (it is $\Z_2$) and the flavor symmetry is $SO(N_f)$.

For example, let us consider an $SO(2)_0\cong U(1)_0$ gauge theory with $N_f=2$ fermions in the fundamental representation, i.e.\ two fermions of charge 1, $\psi^i$. It was suggested in~\XuLXA\ and clarified in~\HsinBLU\ that this theory enjoys self-duality; i.e.\ it is dual to another $U(1)_0$ theory with two fermions $\chi^I$. Unlike the original fermions $\psi^i$, they are labelled by an upper case $I$, because they are acted upon by a different $SU(2)$ global symmetry.  The precise global symmetry, its 't Hooft anomaly, and the deformations of the theory were analyzed in detail in~\BeniniDUS. We will summarize below the facts that are necessary for our discussion.

When the fermions are massless (hence time reversal symmetry is not explicitly broken) and no monopole operators are added to the Lagrangian the model flows to a conformal field theory with an enhanced $O(4)$ symmetry.
This conformal field theory has several notable operators.  First, an $SU(2)$ invariant fermion bilinear in the UV
\eqn\bils{\CO_1=\psi^i\psi_i^{\dagger}}
(with $i=1,2$ labels the two flavors)
flows to an $SO(4)$ singlet in the IR.  Second, the basic monopole operator in the UV $\CM^i$ is in an $SU(2)$ doublet.  $\CM^i$ and its conjugate $\bar \CM^i$ form an $SO(4)$ vector in the IR; i.e.\ the magnetic $U(1)$ is enhanced in the IR to $SU(2)$.  Finally, consider an $SU(2)$ triplet fermion bi-linear
\eqn\bilt{\CO_3^{a,3}=\psi^i\sigma_i^{aj}\psi_j^\dagger}
(the superscript $3$ will be explained shortly).  In the IR it combines with a double-monopole and its conjugate
\eqn\doubm{\eqalign{
&\CM^i\sigma_{ij}^{a}\CM^j=\half(\CO_3^{a,1}+i\CO_3^{a,2}) \cr
&\bar \CM^i\sigma_{ij}^{a}\bar \CM^j=\half(\CO_3^{a,1}-i\CO_3^{a,2}) ~,}}
which are $SU(2)$ triplets, to form a traceless symmetric tensor of $SO(4)$ in the IR.  In terms of $SU(2)\times SU(2)$ representation it transforms as $({\bf 3}, {\bf 3})$ and hence the notation with the two superscripts in $\CO_3^{a,A}$ in~\bilt\ and~\doubm.  The superscript $A$ is distinguished from the superscript $a$ as it denotes a triplet of another $SU(2)$.

In the dual description we have two fermions $\chi^I$ and the flavor $SU(2)$ that mixes them is identified in the IR with the enhanced $SU(2)$ of the original theory.  The singlet \bils\ is identified as $\chi^I\chi_I^\dagger$, the monopole and its conjugate are identified with the $SO(4)$ vector.  And most interestingly, the double-monopole~\doubm\ and the fermion bilinear~\bilt\ are mapped nontrivially
\eqn\tripd{\CO_3^{3,A}= \chi^I\sigma_I^{AJ}\chi_J^\dagger~.}

Our analysis of the phases of the general $SO(N)$ theories can be naturally continued all the way to $SO(2)$ if we properly deform the $SO(2)$ theory such that the UV symmetries are as in the general case.
This can be achieved by adding the charge-two monopole operator and its conjugate, say
\eqn\monop{\CM^i\sigma_{ij}^{3}\CM^j + c.c.=\CO_3^{3,1} }
to the Lagrangian.  (The superscripts $3,1$ where chosen without loss of generality.)  This has the effect of explicitly breaking the flavor $SU(2)$ to $SO(2)$ and the magnetic $U(1)$ to $\Z_2$, such that the UV symmetries are as in a generic $SO(N)$ theory in~\SOsmallk.  In the UV this operator has a very large dimension, but it becomes relevant in the IR.  The duality allows us to analyze its effect in the IR.  This is achieved because in the dual description this operator is a fermion bilinear.  Up to an $SU(2)$ flavor rotation of the dual theory it is~\BeniniDUS
\eqn\dualmon{\CO_3^{0,3} =\chi^1\chi_1^\dagger- \chi^2\chi_2^\dagger~.}

Before adding the monopole operator to the Lagrangian the UV $U(2)$ global symmetry was enhanced in the IR to $SO(4)$ (we suppress discrete factors).  After adding the monopole operator to the Lagrangian the UV symmetry is explicitly broken to $U(1) \times \Z_2 \subset U(2)$ and it is enhanced in the IR to $ U(1)\times U(1)$ (and we again suppress some discrete factors)~\BeniniDUS.

Adding the perturbation \dualmon\ in the dual theory we see that the two fermions are massive and their mass has opposite sign. As a result, when they are integrated out we remain with a $U(1)$ gauge field and no Chern-Simons term. It is dual to a compact scalar and represents the spontaneous breaking of the magnetic $U(1)$ in the dual theory. Using the duality map this corresponds to the spontaneous breaking of the original $SO(2)$ flavor symmetry that remains in the presence of the monopole operator~\BeniniDUS.  This is precisely as implied by~\SOsmallk\  upon setting $N=2$ and $N_f=2$.

We can further consider the phase transitions in~\SOsmallk. The horizontal axis corresponds to a deformation by the singlet mass term~\bils, which is identified as $\chi^I\chi_I^\dagger$. We should add the operator~\bils\ in conjunction with~\monop.  This means that for some value of the singlet mass $\chi^1$ is massless and at another value $\chi^2$ is massless.  Around these two points we have $U(1)_\half$ with $N_f=1$, which is dual to the $O(2)$ Wilson-Fisher theory.  This is exactly as predicted by the dualities in~\SOsmallk.

This conclusion is quite satisfying because of the following reason.  As explained in~\HsinBLU, the self-duality of $U(1)_0$ with $N_f=2$ follows from the web of dualities of~\SeibergGMD, which is supported by a lot of evidence.  So this is not an additional assumption.  Then, the analysis of the deformation by the double-monopole operator, which proceeds via this duality also follows from the same assumption.  This means that at least for $SO(2)$ with $N_f=2$ the intermediate phase scenario that we have been advocating throughout this paper follows logically from the web of dualities of~\SeibergGMD\ without further assumptions!

A similar treatment applies to the $U(1)_\half $ with a $N_f=1$ theory.  This model again has a $U(1)$ magnetic symmetry, which is not present in the general $SO(N)$ case. We thus interpret our phase diagram as being the result of adding a double-monopole to the Lagrangian. Again, we analyze the effect of the double-monopole operator using duality.  The model $SO(2)_{1/2}+\psi$ without monopole operators in the Lagrangian is dual to the  $O(2)$ Wilson-Fisher fixed point with a complex scalar $\Phi$.  The double-monopole perturbation is translated in this theory to $\Phi^2+\bar\Phi^2$.  Now, as we dial the coefficient of the invariant operator $|\Phi|^2$ we encounter one Ising point (where one scalar is massless and the other has positive mass squared), exactly as in~\SOlargek\ if one substitutes $N=2$, $k=\half $, $N_f=1$.

We expect that for large enough $N_f$ this double-monopole operator is irrelevant in the IR fixed point and therefore it cannot split it as above.  Again, this is consistent with our general picture.  Conversely, a microscopic lattice model of these systems might not have the global magnetic $U(1)$ symmetry.  For large enough $N_f$ we recover the $U(1)$ magnetic symmetry in the IR and the IR fixed point is not split.  But for smaller values of $N_f$, where this interaction is relevant in the IR, the fixed point is split and we find some global symmetry breaking there.  This could explain why people who studied microscopic lattice models of this system, which implicitly include the double-monopole operator in the Lagrangian found global symmetry breaking for low values of $N_f$.

\bigskip\bigskip
\noindent{\bf Acknowledgments}

We would like to thank O.~Aharony, A.~Armoni, F.~Benini, D.~Freed, D.~Gaiotto, S.~Snigerov, and E.~Witten for useful discussions.
Z.K. is supported in part by an Israel Science Foundation center for excellence grant and by the I-CORE program of the Planning and Budgeting Committee and the Israel Science Foundation (grant
number 1937/12). Z.K. is also supported by the ERC STG grant 335182 and by the Simons
Foundation grant 488657 (Simons Collaboration on the Non-Perturbative Bootstrap). N.S. was supported in part by DOE grant DE-SC0009988.

\listrefs
\end